\documentclass[
 reprint,
 amsmath,amssymb,
 aps,
 prd,
]{revtex4-2}

\usepackage{graphicx}
\usepackage{dcolumn}
\usepackage{dsfont}
\usepackage{braket}
\usepackage{subcaption}
\usepackage{xspace}
\usepackage{hyperref} 
\usepackage{tikz}

\graphicspath{
    {./figs}%
    {./tikz}%
    }

\begin{document}

\title{Wigner rotations for cascade reactions}

\author{Kai Habermann}
\affiliation{HISKP, University of Bonn, Nussallee 14-16,  Bonn, Germany}
\email{kai.sebastian.habermann@cern.ch}

\author{Mikhail Mikhasenko}
\affiliation{Ruhr University Bochum, Universitätsstraße 150, Bochum, Germany}
\email{mikhail.mikhasenko@cern.ch}

\date{\today}

\newcommand{\reftext}{\ensuremath{\mathrm{\frak{o}}}}
\newcommand{\chainind}{\ensuremath{\frak{c}}\xspace}
\newcommand{\chainone}{\ensuremath{\frak{1}}\xspace} 
\newcommand{\chaintwo}{\ensuremath{\frak{2}}\xspace}
\newcommand{\chainthree}{\ensuremath{\frak{3}}\xspace}

\newcommand{\rftext}{\ensuremath{\mathrm{r.f.}}}
\newcommand{\minusphi}{\textit{minus\_phi}\xspace}
\newcommand{\helicity}{\textit{helicity}\xspace}
\newcommand{\canonical}{\textit{canonical}\xspace}
\newcommand{\particletwo}{\textit{particle-two}\xspace}
\newcommand{\collective}[1]{\ensuremath{#1_{1},\dots,#1_n}\xspace}
\newcommand{\WignerAll}[1]{\ensuremath{W_{\collective{#1'};\collective{#1}}^{\chainind(\reftext)}}}
\newcommand{\CG}[6]{C_{#2, #4}^{#1, #3 | #5}}
\newcommand{\ie}{\textit{i.e.}\xspace}
\newcommand{\eg}{\textit{e.g.}\xspace}

\newcommand{\mother}{{\ensuremath{{\kappa}}}\xspace}
\newcommand{\done}{{\ensuremath{\kappa{[1]}}}\xspace}
\newcommand{\dtwo}{{\ensuremath{\kappa{[2]}}}\xspace}

\newcommand{\intdecgen}[2]{\ensuremath{#1}\xspace}
\newcommand{\nodeind}{{\ensuremath{\kappa}}\xspace}

\newcommand{\interntext}{\text{internal}}

\newcommand{\Hnode}[1]{\ensuremath{H_{#1}^{\nodeind}}\xspace}
\newcommand{\hnodegen}[2]{\ensuremath{h_{#1}^{#2}}\xspace}
\newcommand{\hnode}[1]{\ensuremath{\hnodegen{#1}{\nodeind}}\xspace}


\newcommand{\SLtwoC}{\ensuremath{\mathrm{SL}(2,\mathbb{C})}\xspace}
\newcommand{\SOtreeone}{\ensuremath{\mathrm{SO}^+(3,1)}\xspace}

\begin{abstract}
    Parametrization of cascading hadronic reactions is a central tool in hadron spectroscopy for modeling matrix elements and extracting parameters of hadronic states.
    Implementing the helicity formalism consistently presents challenges,
    particularly for particles with spin, due to the need to match spin states of final-state particles, an operation known as the Wigner rotation.
    This paper discusses these challenges in detail and offers solutions,
    including a practical method for implementation.
    Equipped with a general algorithm for computing Wigner rotations,
    we extend the studies to alternative amplitude formulations, the minus-phi and canonical conventions.
\end{abstract}

\maketitle

\section{Introduction}\label{sec:introduction}

The analysis of multibody decays is an essential method for studying hadronic resonances,
leading to a significant increase in the number of observed hadronic states,
some of which challenge the constituent quark model~\cite{Karliner:2017qhf,Klempt:2009pi,Richard:2016eis,Guo:2017jvc,Esposito:2016noz}.
Among these are pentaquarks observed in the
$\Lambda_b^0 \rightarrow J/\psi p K^-$ decays~\cite{LHCb:2015yax,LHCb:2019kea}
and hidden-charm strange states in $B^+ \rightarrow J/\psi \phi K^+$ decays~\cite{LHCb:2016axx,LHCb:2016nsl,LHCb:2021uow},
as well as charged tetraquarks in \mbox{$e^+ e^- \to \pi^- \pi^+ J/\psi$} processes~\cite{BESIII:2020oph}
and $B^0 \rightarrow \psi K^+ \pi^-$ decays~\cite{PhysRevD.88.074026}.
With more data of increasingly better quality,
many observations of hadronic states are driven by amplitude analysis,
a technique that utilizes angular correlations and Lorentz group properties to constrain the spin and parity of particles.
Insights into the dynamics within individual subsystems are obtained by modeling the matrix element of the reaction.
However, examining the mass spectrum of the subsystem is not sufficient since the cross-channel contributions -- referring to competing coherent processes in different combinations of particles where an intermediate resonance occurs -- create a background. Amplitude analysis is the only known systematic method of distinguishing these contributions.

Modeling dynamics of a multibody final state is a complex process.
One common approach is the \textit{isobar model}~\cite{Herndon:1973yn,Hansen:1973gb}, where the multibody decay is represented as a series of two-body decays assuming intermediate resonances.
The notion of the isobar model is widely used in the literature as a vague term, sometimes implying that Breit-Wigner parametrization is used to model resonances. More advanced approaches such as the K-matrix~\cite{Aitchison:1972ay,ParticleDataGroup:2022pth} are also commonly used~\cite{FOCUS:2003tdy,LHCb:2024cwp}, with the Khuri-Treiman framework~\cite{Khuri:1960zz,Aitchison:1979fj,Pasquier:1968zz} being one of the most sophisticated ways of describing resonance lineshapes in three-body decays.
It accounts for final-state interactions by replacing the Breit-Wigner parametrization with functions derived from solving integral equations specific to each process. Phenomenological modeling using effective-field-theory methods is also used to derive a matrix element for sequential decays, \eg see~\cite{Altmannshofer:2008dz,LHCb:2024onj}.
All such models, simple or advanced, can be mapped into the angular-function basis described here, with differences arising only from the choice of resonance lineshapes.

In this paper, we refer to the modeling method as the \textit{cascade-decay} parametrization,
meaning a general framework that covers all categories above
by focusing on angular variables while leaving the energy dependence out of scope.
One finds an extended discussion on the impact of different formalisms for energy dependence in Refs.~\cite{Filippini:1995yc,Chung:2007nn,JPAC:2017vtd,JPAC:2018dfc,Chen:2017gtx}.
We assume that the amplitude is written as a sum of terms corresponding to distinct topologies, with each part seen as a truncated series of partial waves.
In this paper we focus on decays for simplicity,
although the methods and arguments presented here are equally applicable to scattering amplitudes
modeled using the partial-wave expansion.

Lorentz group properties play a central role in describing cascade reactions.
The particle spin states necessitate basis changes, which are performed using Wigner D-functions, a representation of the rotation group.
It is widely understood that in cascade decays, a set of Wigner matrices appears in the description of the sequential decays.
Less straightforwardly, another set of Wigner matrices is required to ensure consistent treatment of spin quantization axes across various coherent processes contributing to the amplitude.
These alignment rotations are known as the \textit{Wigner rotations}.
This paper aims for an extended discussion on the computation and application of the Wigner rotations within the parametrization of cascade reactions.
Although several papers~\cite{Chen:2017gtx,Li:2022qff,Marangotto:2019ucc} have discussed how decay amplitudes are constructed using different formalisms, the computation of Wigner rotations, which are essential in any formalism, has often been set aside.
Analytic expressions for Wigner rotations in simple aligned three-body kinematics are derived in Ref.~\cite{JPAC:2019ufm}. Reference~\cite{Wang:2020giv} uses \SLtwoC group elements and clever particle ordering to circumvent complications in Wigner rotations.
To our knowledge, no existing reference details the computation of Wigner rotations in the context of general cascade amplitudes.
These alignment factors are integral to every amplitude analysis, including those that discover and study exotic hadronic configurations in experiments worldwide.

A comment is due that tensor formulations~\cite{Anisovich:2006bc,Zemach:1965ycj,Chung:1997jn} do not use spin states explicitly, and therefore, are exempt from the need of Wigner rotations.
A consistency of quantization axes is built in automatically by using the same tensor structure for decay chains.

The structure of this paper is as follows: we begin with a discussion of kinematics in Sec.~\ref{sec:kinematics}.
This is followed by Sec.~\ref{sec:parameters-of-lorentz-transformations} with an examination of the Lorentz group, its representations, and our approach to implement the group operations.
Next, we review the helicity parametrization of cascade reactions and introduce the Wigner rotations in Sec.~\ref{sec:parametrization-of-cascade-reactions}.
Section~\ref{sec:discussion-on-application} is dedicated to
a discussion of subtleties and possible caveats involved in amplitude construction.
Finally, in Sec.~\ref{sec:other-formulations} we discuss alternative conventions and investigate the impact of the choice of convention on the Wigner rotations. We then conclude with a summary of our findings.

\section{Kinematics} \label{sec:kinematics}

The phase space for $n$ particles can be parametrized using variables of $n-1$ two-particle systems in a recursive phase-space representation~\cite{Byckling:1971vca}. Each two-particle node in the splitting graph is described by three variables ($m, \theta, \phi$). With the mass of the first node fixed to the total mass, the phase-space domain is defined by $3n-4$ kinematic variables.

The choice of kinematic variables is intrinsically linked to the form of the decay matrix element. Specific decay models often necessitate particular sets of variables. It is common practice to use distinct sets of variables when calculating different coherent parts of the amplitude. However, employing different sets of variables within the same amplitude, while being necessary, leads to severe complications, which will be addressed in detail below.

To facilitate notation, we use \nodeind to label nodes in the decay graph. Internal nodes correspond to a decay \mbox{$\mother\to\done,\dtwo$}, where \mother represents the decaying system, and \done and \dtwo are the ordered node labels for the decay products. Also, each node, \mother, \done, and \dtwo, is associated with either intermediate particle subsystems or a final-state particle. An example using bracket notation (see also Appendix~\ref{sec:example-amplitude-explicitly}) further illustrates the node labeling:
\begin{center}
    \includegraphics{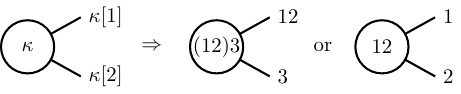}
\end{center}

In the rest frame of $\mother$, the solid angle $\varOmega_\nodeind$ is represented by the pair $(\theta_\nodeind, \phi_\nodeind)$, which are the spherical angles of the first particle in the daughter pair. These angles can be computed as follows:
\begin{align}
    \label{eq:theta}
    \theta_\nodeind = & \cos^{-1}\left( \frac{p_z}{|\vec{p}\,|} \right) \quad\text{with}\quad p=p_\done\,, \\
    \label{eq:phi}
    \phi_\nodeind =   & \tan^{-1}\left( p_y, p_x\right) \quad \text{with}\quad p=p_\done\,.
\end{align}
It is to note, that both particle $\done$ and particle $\dtwo$ may themselves decay.
%
%
In the \textit{helicity} convention, the rest frame of $\done$ or $\dtwo$ is approached by
a set of active transformations, $\Lambda_{\mother\leftarrow \done}^{-1}$ or $\Lambda_{\mother \leftarrow \dtwo}^{-1}$, respectively.
\begin{align} \nonumber
    \Lambda^{(\text{hel.})}_{\mother \leftarrow \done} & = R(\phi_\nodeind, \theta_\nodeind, 0) B_z(\gamma_\nodeind^\done)\,,          \\ \label{eq:hel}
    \Lambda^{(\text{hel.})}_{\mother \leftarrow \dtwo} & = R(\phi_\nodeind, \theta_\nodeind, 0) R_y(\pi) B_z(\gamma_\nodeind^\dtwo)\,,
\end{align}
where \mbox{$R(\phi_\nodeind, \theta_\nodeind, 0) = R_z(\phi_\nodeind) R_y(\theta_\nodeind)$}, and $B_z$,
$R_y$, and $R_z$ are operators from the Lorentz group acting on the particle states, $\gamma_\nodeind^\done$ and $\gamma_\nodeind^\dtwo$ denote the Lorentz factors of particles \done and \dtwo, respectively, in the $\mother$ frame.
\begin{figure}[ht]
    \includegraphics{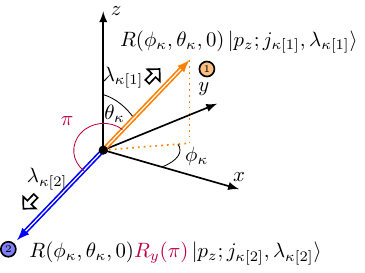}
    \caption{
        Definition of the helicity angles ($\phi_\nodeind,\theta_\nodeind$) in the decaying particle rest frame.
        The particle-one and particle-two momentum vectors are indicated with a  circled index.
        The shaped arrows indicate the quantization axes of the particle spins.
        Formulae list the definitions of the two one-particle states as used in the \helicity convention.
    }
    \label{fig:coordinate-system}
\end{figure}
Explicit representations of the transformation matrices for the spin states are provided in the Appendix~\ref{sec:matrices-for-lorentz-transformations}.

For a particle with a four-momentum $p$, and spin $j$, a helicity $\lambda$ is defined as a projection of its spin onto the direction of flight.
A helicity state can be constructed by applying active boost and rotation transformations to the rest-frame state quantized along the $z$~axis.
\begin{align} \label{eq:hel.state}
    \ket{p; j,\lambda} = R B_z \ket{0; j,\lambda}\,.
\end{align}
The helicity state is not invariant under boosts, so it is important to know in which frame it is defined.
It also receives a phase factor under rotations, despite the fact that the helicity value stays the same.

The helicity states of the particles $\done$ and $\dtwo$ in the rest frame of the mother particle $\mother$,
are given by $\Lambda^{(\text{hel.})}_{\mother \leftarrow \done} \ket{0;j_{\done},\lambda_{\done}}$, and $\Lambda^{(\text{hel.})}_{\mother \leftarrow \dtwo} \ket{0;j_{\dtwo},\lambda_{\dtwo}}$, respectively.
The transformation for the second particle $\dtwo$ in Eq.~\eqref{eq:hel} is different from the naive expression, \mbox{$R(\pi+\phi_\nodeind,\pi-\theta_\nodeind,0)$}.
A critical need to use the same rotation \mbox{$R(\phi_\nodeind, \theta_\nodeind,0)$} for both daughter particles is further discussed in Sec.~\ref{sec:parametrization-of-cascade-reactions}, when introducing helicity couplings.

\section{Parameters of Lorentz transformations}
\label{sec:parameters-of-lorentz-transformations}

An arbitrary Lorentz transformation can be represented in terms of a combination of rotations and a single boost:
\begin{align}\label{eq:arb.lorentz}
    \Lambda = R(\phi, \theta, 0) B_z(\gamma) R(\phi_\rftext, \theta_\rftext, \psi_\rftext)\,,
\end{align}
where the subscript $\rftext$, for rest frame, indicates that the rotation is applied before the boost, and
\mbox{$R(\phi, \theta, \psi) = R_z(\phi) R_y(\theta) R_z(\psi)$}.
In this parametrization, the Lorentz transformation is fully described by the six real numbers $\phi$, $\theta$, $\gamma$, $\phi_\rftext$, $\theta_\rftext$, and $\psi_\rftext$.
The ranges of the angles are as follows: $\phi$ and $\psi_\rftext$ in $[-\pi, \pi]$, $\theta$ and $\theta_\rftext$ in $[0, \pi]$, and $\phi_\rftext$ in $[-\pi, 3\pi]$.
The necessity of the extended range of the $\phi_\rftext$ angle becomes clear when considering odd-dimensional representations of the Lorentz group corresponding to
half-integer spin.
For these matrices, a $2\pi$ shift in the $\phi_\rftext$ angle results in
a sign flip for the matrix $\Lambda$.

The process of decoding an arbitrary sequence of boosts and rotations into the form of Eq.~\eqref{eq:arb.lorentz} requires determining
the set of six parameters $\phi$, $\theta$, $\gamma$, $\phi_\rftext$, $\theta_\rftext$, and $\psi_\rftext$.
We use a trick to do that: a combination of two representations of the Lorentz group \SLtwoC, acting on Dirac spinors, and \SOtreeone, acting on four-momenta, is utilized.
It is easy to see from Eq.~\eqref{eq:arb.lorentz} that the parameters $\theta$, $\phi$, and $\gamma$ can be obtained from the effect of the transformation on a four-vector with zero spatial components.
With these values determined, we compute the rest-frame rotation part of the matrix as
\begin{equation} 
    R_\rftext = B_z^{-1}(\gamma) R_y^{-1}(\theta) R_z^{-1}(\phi) \,\Lambda
\end{equation}
The angles $\phi_\rftext$, $\theta_\rftext$, and $\psi_\rftext$ are obtained from the \SLtwoC representation in the case of a pure rotation.
The procedure is detailed in Appendix~\ref{sec:decoding-parameters-of-lorentz-transformation}.

\section{Parametrization of Cascade Reactions} \label{sec:parametrization-of-cascade-reactions}

We formally define the matrix element describing an $n$-body decay
from the initial state labeled $0$ to the final state with particles labeled $1, \ldots, n$ by the expression,
\begin{align} \nonumber
    \bra{\text{final}}T\ket{\text{initial}} & =  \mathcal{A}_{\lambda_0;\collective{\lambda}}(\tau_\reftext) \\ \label{eq:A.def}
                                            & \qquad \, \times (2\pi)^4 \delta^4(p_0 - \sum_i^n p_i),
\end{align}
where the initial state is a canonical state in the center-of-momentum frame, $\ket{\text{initial}} = \ket{0; j_0,\lambda_0}$, and
the final state is a direct product of helicity states for all final-state particles, $\ket{\text{final}} = \prod_i \ket{p_i,\lambda_i}$.
The transition amplitude $\mathcal{A}_{\lambda_0;\collective{\lambda}}(\tau_\reftext)$ is a complex function, which is continuous in kinematic variables, collectively named $\tau_\reftext$, and is dependent on discrete indices of particle-spin projections $\lambda_0$ and $\collective{\lambda}$. We use a lower Fraktur index \reftext for the reference set of kinematic variable.
The cascade parametrization models the dynamics of the decay as a sum of coherent processes, each representing sequential decays with intermediate resonances:
\begin{align}\nonumber
    \mathcal{A}_{\lambda_0;\collective{\lambda}}(\tau_\reftext) & = \sum_{\chainind}\sum_{\collective{\lambda'}}  \mathcal{A}^\chainind_{\lambda_0;\collective{\lambda'}}(\tau_\chainind) \\ \label{eq:Afull}
                                                                & \qquad \times \WignerAll{\lambda}(\tau_\chainind|\tau_\reftext) \,,
\end{align}
where the Fraktur index $\chainind$ numbers the decay chains,
and $\tau_\chainind$ is a set of kinematic variables specific to the chain $\chainind$.
The primed helicity indices refer to spin projections with the quantization axes specific to the chain $\chainind$.
The term $\WignerAll{\lambda}(\tau_\chainind|\tau_\reftext)$ implements Wigner rotations to ensure a consistent treatment of the quantization axes across different chains. This term is the central object of discussion for this paper and will be further elaborated on in the following paragraphs.
\begin{figure*}[htb]
    \centering
    \includegraphics{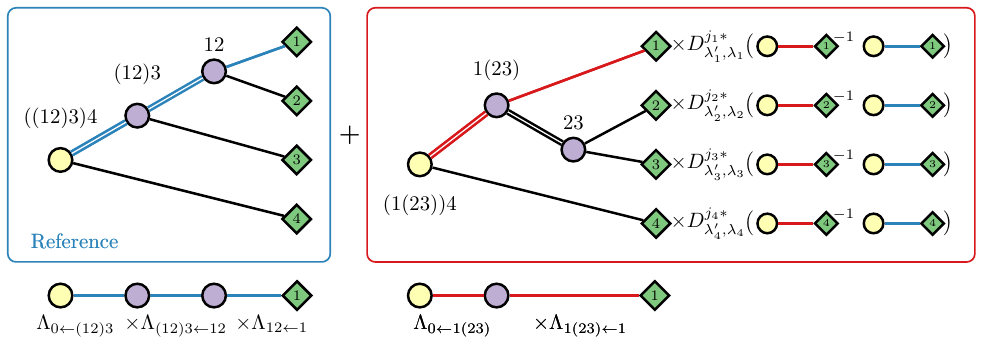}
    \caption{
        A sketch illustrating computations in Eq.~\eqref{eq:A.def} using a four-body decay example.
        Each node in the decay graph corresponds to the rest frame of a subsystem of particles, identified by its label. The yellow circular node represents the root node, corresponding to the center-of-momentum frame, while the diamond-shaped nodes represent the rest frames of the final-state particles.
        The paths through the graph are identified with sequences of Lorentz transformations connecting rest frames associated with the nodes. The blue color refers to the first left topology, used as the reference, and the red color refers a second right topology.
        The blue and red paths below the figure depict two distinct ways to reach the final-state particle 1 from the root node, corresponding to the left and right topologies, respectively.
        Equations~\eqref{eq:wigner.product} and \eqref{eq:L.rot} are illustrated in the diagram on the right, where the arguments of the alignment rotations are computed using
        distinct paths from the root node to the final-state nodes for each particle.
        The complete amplitude expression for this example is detailed in Appendix~\ref{sec:example-amplitude-explicitly}.
    }
    \label{fig:wigner.rotation}
\end{figure*}

The chain amplitude $\mathcal{A}^{\chainind}_{\lambda_0;\collective{\lambda^\prime}}(\tau_\chainind)$ represents the amplitude
of a single decay chain connecting the initial state with the final state.
A \textit{cascade topology} or \textit{decay graph} is a tree of two-body transitions.
Within a decay graph \textit{nodes} can be labeled in various ways; we index them with \nodeind.
Each node corresponds to the center-of-momentum frame of the decay pair as approached from the parent node.
A decay topology, accompanied by quantum numbers for all internal nodes, is referred to as a \textit{decay chain}.
The chain amplitude is obtained by multiplying the two-body amplitudes associated with each node of the decay chain.
\begin{align} \nonumber
    \mathcal{A}^{\chainind}_{\lambda_0;\collective{\lambda'}}(\tau_\chainind) & = \sum_{\lambda_\interntext'\dots}\prod_{\nodeind}
    D^{j_{\mother}*}_{\lambda_{\mother}',\lambda_{\done}'-\lambda_{\dtwo}'}(\varOmega_\nodeind^\chainind) X_\nodeind(\mu_{\mother})                                               \\ \label{eq:Achain}
                                                                              & \quad \times \Hnode{\lambda_{\done}', \lambda_{\dtwo}'}(\mu_{\mother},\mu_{\done},\mu_{\dtwo})\,,
\end{align}
where, $j_{\mother}$ is the spin of $\mother$, $D$ is a Wigner D-function.
In the \helicity conventions, $D^{j}_{\rho,\eta}(\varOmega) = D^{j}_{\rho,\eta}(\phi, \theta, 0)$.
The product index \nodeind runs over all \textit{nodes} of the decay graph, where each node \nodeind describes a decay $\mother\to\done,\dtwo$.
States in the intermediate propagators (all internal nodes except the root node) have their helicity values appearing twice: as $\lambda_{\done}'$ or $\lambda_{\dtwo}'$ for the production node and as $\lambda_{\mother}'$ if the state decays further. These values must be the same, and their equality is ensured by transformations between frames. Additionally, these values are summed over as indicated by $\lambda_\interntext'$, as these internal helicities are not observable quantities.
The $X_\nodeind(\mu_\mother)$ parametrizes the lineshape function and depends on the mass of the system \mother, $\mu_\mother = \sqrt{p_\mother^2}$,
and \mbox{$\Hnode{\lambda_{\done}', \lambda_{\dtwo}'}$} is the helicity vertex function, that might depend on all masses related to the node \nodeind.
In practice, it's more convenient to deal with the helicity couplings in the \particletwo convention, denoted by $\hnode{\lambda_{\done}, \lambda_{\dtwo}}$. The couplings $H$ and $h$ differ by a phase,
\begin{align} \label{eq:particle.two.phase}
    \Hnode{\lambda_{\done}, \lambda_{\dtwo}} = (-1)^{j_{\dtwo}-\lambda_{\dtwo}} \hnode{\lambda_{\done}, \lambda_{\dtwo}}\,,
\end{align}
that arises from the definition of the particle-two state~\cite{Martin:1970hmp,Richman:1984gh}.
The couplings $h$ have convenient transformation properties under parity,
\begin{equation}
    \hnode{-\lambda_{\done}, -\lambda_{\dtwo}} = (-1)^{j_{\mother}-j_{\done}-j_{\dtwo}} P_{\mother}P_{\done}P_{\dtwo} \,\hnode{\lambda_{\done}, \lambda_{\dtwo}}\,.
\end{equation}
where $P_{\done}$, $P_{\dtwo}$, and $P_{\mother}$ are parities of the daughter particles and the mother particle for the node \nodeind.
It is mapped to spin-orbit vertex $\hnode{l_\nodeind s_\nodeind}$ couplings using Clebsch-Gordan coefficients,
\begin{align} \label{eq:ls}
     & \hnode{\lambda_{\done}, \lambda_{\dtwo}} =  \sum_{l_\nodeind s_\nodeind} \hnode{l_\nodeind s_\nodeind} \,
    \CG{j_{\done}}{\lambda_{\done}}{j_{\dtwo}}{-\lambda_{\dtwo}}{s_\nodeind}{\Delta \lambda_{\nodeind}} \,
    \CG{l_\nodeind}{0}{s_\nodeind}{\Delta \lambda_{\nodeind}}{j_{\mother}}{\Delta \lambda_{\nodeind}}\,,
\end{align}
with $\Delta \lambda_{\nodeind} = \lambda_{\done}-\lambda_{\dtwo}$ and the indices $l_\nodeind$ and $s_\nodeind$ going over all possible combinations of ordinal angular momentum and spin of the $(\done,\dtwo)$ pair.
Indices of the Clebsch-Gordan coefficients map into the standard notations~\cite{ParticleDataGroup:2022pth} as follows,
$$
    \CG{j_1}{\lambda_1}{j_2}{\lambda_2}{j_3}{\lambda_3} = \left\langle j_1,\lambda_1; j_2,\lambda_2| j_3,\lambda_1+\lambda_2\right\rangle\,.
$$
Note the minus sign for particle-two helicity in Eq.\eqref{eq:ls}.
It appears because particle one is used as the reference, as shown in Fig.\ref{fig:coordinate-system}.

Wigner rotations arise in Eq.~\eqref{eq:Afull} due to the mismatch of quantization axes between different decay chains.
One finds the primary cause of the Wigner rotations when examining what helicities are listed as arguments of $\mathcal{A}^{\chainind}_{\lambda_0;\collective{\lambda'}}$ in Eq.~\eqref{eq:Achain}. Building the decay amplitude implies moving with Lorentz transformations through the decay graph from the initial node toward decay products.
In this procedure, the helicity value of every particle appears once, and from a single node, which corresponds to a particular rest frame of the reaction subsystem. This frame defines the particle helicity based on transformations from Eq.~\eqref{eq:hel}. Hence, the particle helicity values $\collective{\lambda'}$ in argument of $\mathcal{A}^{\chainind}_{\lambda_0;\collective{\lambda'}}$ are defined in various frames within the same decay chain. Moreover, which frames define the helicities varies between different decay chains.

At this point, it comes with no surprise that aligning the quantization axes between different chains is crucial.
It is practical to select one of the decay chains as a \textit{reference chain} and apply the Wigner rotations to all other chains to align the quantization axes with the reference.
Let us compare a helicity state in a particular moving frame, defined by Eq.~\eqref{eq:hel.state}
with a reference transformation where the path from the rest frame to the moving frame
is given by a general transformation $\Lambda$.
Using Eq.~\eqref{eq:arb.lorentz}, one finds,
\begin{align} \nonumber
    \Lambda \ket{0; j,\lambda} & = R B_z R_\rftext \ket{0; j,\lambda}                                         \\ \label{eq:p.ket.rf}
                               & = \sum_{\lambda'} D^{j}_{\lambda',\lambda}(R_\rftext) \ket{p; j,\lambda'}\,.
\end{align}
The state vectors in Eq.~\eqref{eq:hel.state} and Eq.~\eqref{eq:p.ket.rf} share the same momentum value. The rotation $R_\rftext$ in the particle's rest frame affects only the helicity quantum number. This rest-frame rotation accounts for the difference between two Lorentz transformations that connect the moving frame to the rest frame of a particle.
In the context of decay amplitude construction, the general transformation $\Lambda$ is associated with the reference topology. We denote the rest-frame mismatch rotation between the reference chain $\reftext$ and the chain $\chainind$
for the final-state particle $i$ as $R_{\chainind(\reftext)}^i$.
The coefficients $D_{\lambda',\lambda}^j$ for transforming the final state appear conjugated in Eq.~\eqref{eq:Afull} according to Eq.~\eqref{eq:A.def}.
Namely,
\begin{equation} \label{eq:wigner.product}
    \WignerAll{\lambda} = \prod_{i=1}^n D_{\lambda_i',\lambda_i}^{j_i*}(
    R_{\chainind(\reftext)}^i)\,.
\end{equation}
where
$D^{j}_{\lambda',\lambda}(R_{\chainind(\reftext)}^i)$ is a short notation for \mbox{$D^{j}_{\lambda',\lambda}(\phi_{\chainind(\reftext)}^i,\theta_{\chainind(\reftext)}^i,\psi_{\chainind(\reftext)}^i)$} referring to the angles of the rotation $R_{\chainind(\reftext)}^i$;
$R_{\chainind(\reftext)}^i$ is the \textit{Wigner rotation} that relates the chain $\chainind$ to a reference chain, and the index $i$ numbers the final-state particles.

To calculate the relative rotation between the frames reached by traversing two different topologies,
we compare a series of transformations between different chains.
\begin{align} \label{eq:L.rot}
    R_{\chainind(\reftext)}^i = &
    \left(\prod_a^{0 \rightarrow i} \Lambda_{a \leftarrow a_\text{next}}^{\chainind} \right)^{-1}
    \times
    \left( \prod_b^{0 \rightarrow i} \Lambda_{b \leftarrow b_\text{next}}^{\reftext} \right)\,,
\end{align}
where the transformations $\Lambda_{a \leftarrow a_\text{next}}^{\chainind}$ are defined in Eq.~\eqref{eq:hel}.
The first matrix product is built from left to right while traversing all internal nodes from the initial state $0$ to the final state $i$,
following the path set by chain $\chainind$. The second product follows the same procedure for the reference topology.
Figure~\ref{fig:wigner.rotation} demonstrates the construction of these matrix products.
The resulting transformation $R_{\chainind(\reftext)}^i$ rotates from the rest frame of particle $i$ reached by traversing the reference topology to the rest frame obtained by following topology $\chainind$.
A sketch of this construction for a four-body decay is shown in Fig.~\ref{fig:wigner.rotation}, and a detailed expression of the example amplitude is provided in Appendix~\ref{sec:example-amplitude-explicitly}.

\section{Discussion on Application}
\label{sec:discussion-on-application}

We begin by discussing the general principle of rotation factorization in $n$-body final states.
As pointed in Ref.~\cite{JPAC:2019ufm}, for any system with a fixed total spin $j_0$, the overall rotation of the final-state momenta can be factored out,
simplifying the analysis of the remaining dynamics.
It reads,
\begin{align} \label{eq:dpd}
    \mathcal{A}_{\lambda_0;\collective{\lambda}}(\tau) & = \sum_{\lambda_0^\prime}
    D^{j_0*}_{\lambda_0,\lambda_0'}(\alpha, \beta, \gamma) \mathcal{A}^{\text{aligned}}_{\lambda_0';\collective{\lambda}}(\tau_{\text{align}}),
\end{align}
where $(\alpha, \beta, \gamma)$ are the Euler angles of the overall rotation and $\tau_{\text{align}}$ encapsulates $3n - 7$ kinematic variables of the aligned reaction.
These angles can be chosen from a reference decay chain as
$\phi_{A}^\reftext$, $\theta_{A}^\reftext$, and $\phi_{B}^\reftext$ for $\alpha$, $\beta$, and $\gamma$, respectively,
where $A$ is a root node, and $B$ is an internal node connected to $A$.
The factorization is not used explicitly in Eq.~\eqref{eq:Afull},
but it comes as a consequence of the construction if implemented correctly.

The general case of a three-body decay is complex enough to manifest a subtle feature of an additional $2\pi$ phase that cannot be captured with the rotation group of particle four-momenta, \SOtreeone.
The labels 1, 2, 3 are used for three final-state particles, and $0$ is used for the decaying particle.
We consider three decay chains distinguished by their topologies $(23)1$, $(31)2$, and $(12)3$, labeled by the spectator index in Fraktur as \chainone, \chaintwo, and \chainthree, respectively.
The chain \chainone is used as the reference chain.
The kinematics is parametrized by $\tau_\chainone = (\phi_{(23)1}, \theta_{(23)1}, m_{23}, \phi_{23}, \theta_{23})$.
We omit the upper Fraktur indices for three-body kinematics to simplify notations since each index uniquely relates to a topology.
Comparing this case to Eq.~\eqref{eq:dpd}, the overall rotation is given by $R(\phi_{(23)1}, \theta_{(23)1}, \phi_{23})$.

\subsection{Kinematics mismatch}
\label{ssec:kinematics-mismatch}

\begin{figure*}[t]
    \centering
    \begin{subfigure}[t]{0.47\linewidth}
        \centering
        \includegraphics[width=\textwidth]{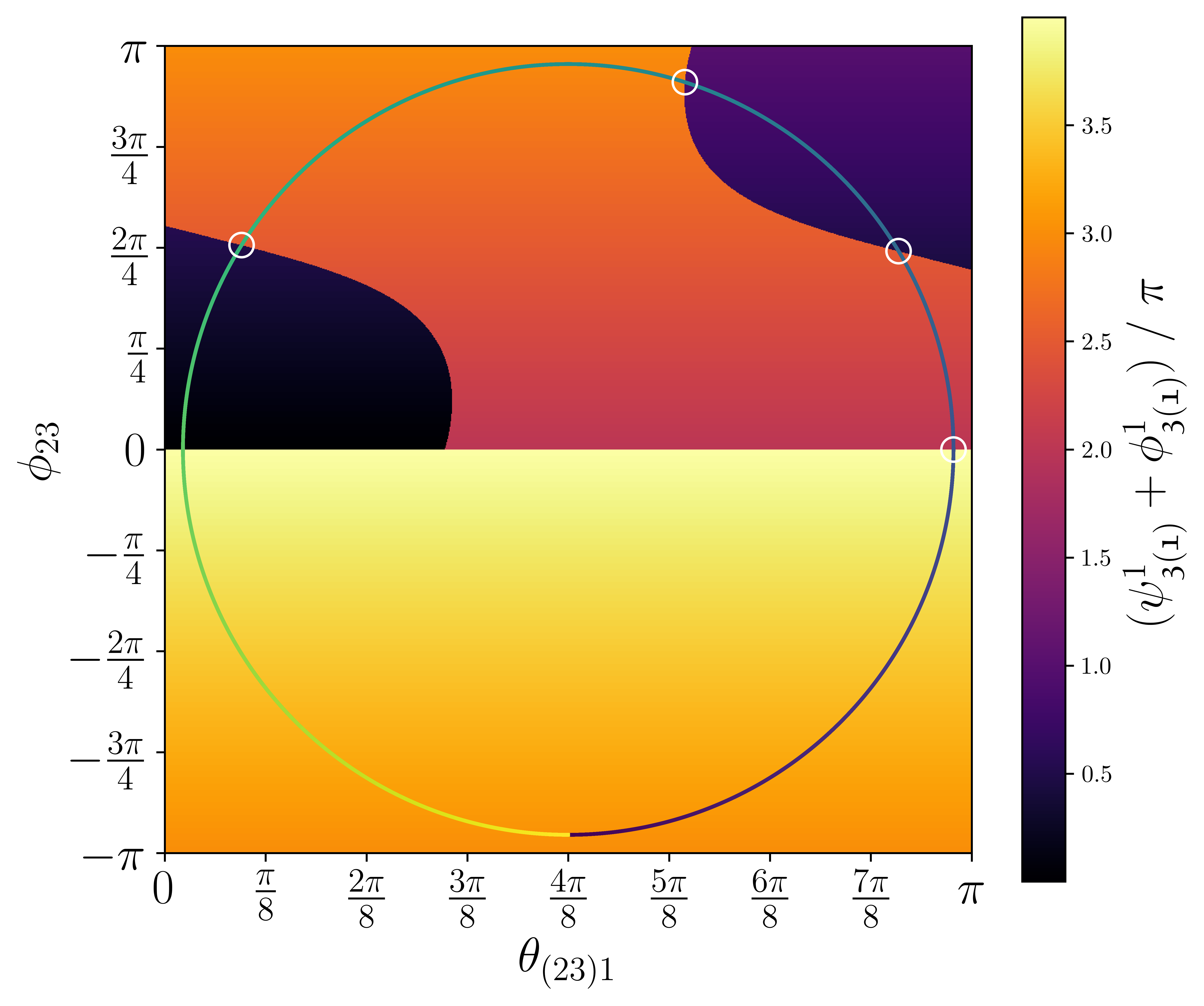}
        \caption{Azimuthal part of the Wigner rotation}
        \label{fig:helicity}
    \end{subfigure}
    \hfill
    \begin{subfigure}[t]{0.47\linewidth}
        \centering
        \includegraphics[width=0.95\textwidth]{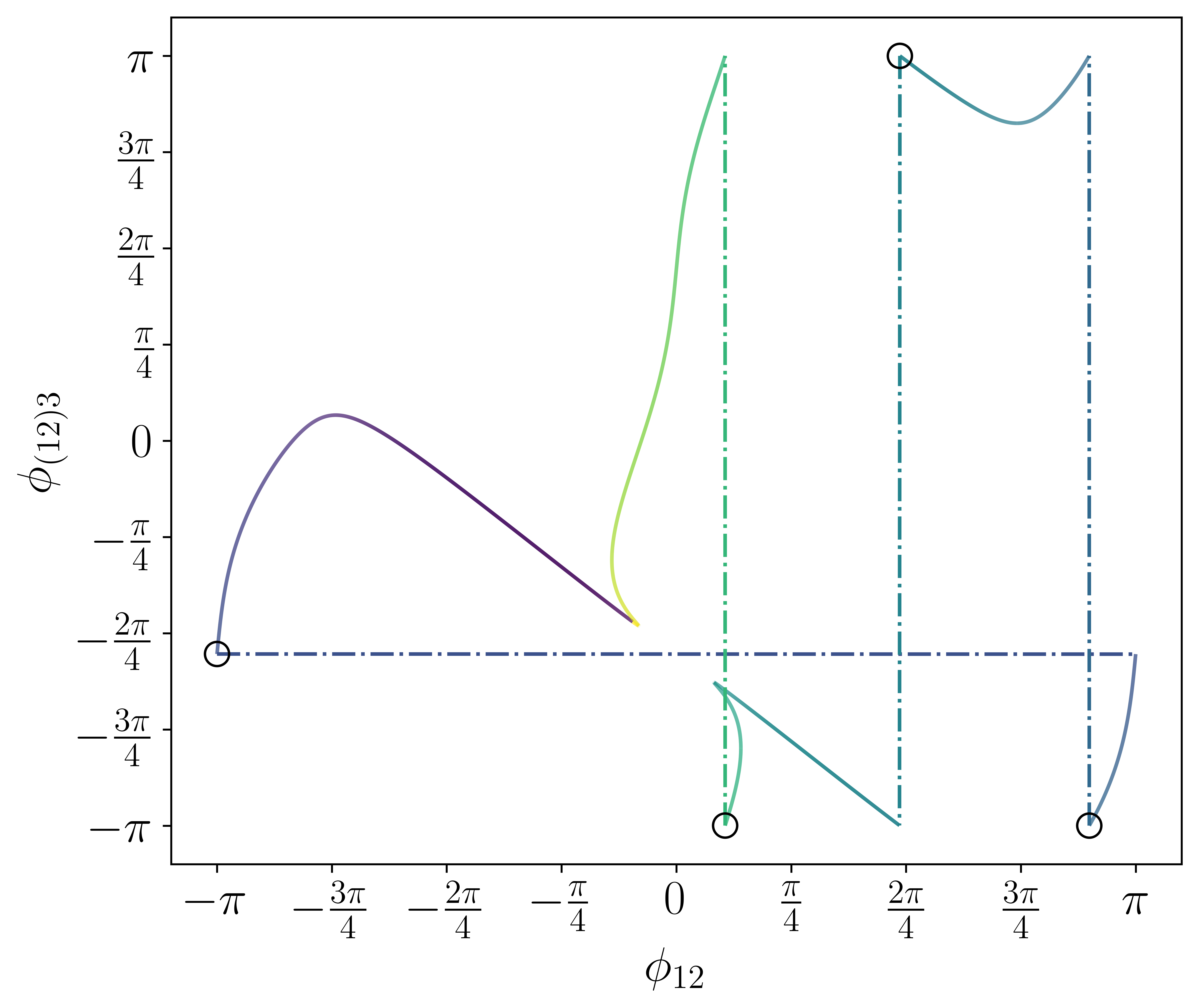}
        \caption{Azimuthal helicity angles along the circular path}
        \label{fig:helicity_jumps}
    \end{subfigure}
    \caption{
        The origin of $2\pi$ discontinuity in azimuthal Wigner rotations is demonstrated using a three-body decay as an example.
        (a) Azimuthal part of Wigner rotation, $\psi_{\chainthree(\chainone)}^1 + \phi_{\chainthree(\chainone)}^1$, for particle~$1$ in the chain $0 \to (12)3$, using the chain $0 \to (23)1$ as a reference computed with the \textit{helicity} convention as a function of kinematic variables is depicted by the color. The $x$, and $y$-axes show the rotation angles $\theta_{(23)1}$ and $\phi_{(23)1}$, respectively, that are used to generate the four-vectors from the aligned configuration, $\phi_{(23)1}$ set to $1.4$.
        The colored line shows a continuous trajectory in the phase space.
        The white circles pinpoint the $2\pi$ discontinuities~\ref{fig:helicity_jumps}.
        (b) Azimuthal helicity angles: the root node angle, $\phi_{3(12)}$ and subchannel 12 angle, $\phi_{12}$
        along the circular path marked on the figure (a). The marked points locate the $2 \pi$ jumps. They can be related to the left plot using the color of the circular path.
    }
    \label{fig:origin.of.2pi}
\end{figure*}

Figure~\ref{fig:helicity} shows the value of the azimuthal part of the Wigner rotation $R_{\chainthree(\chainone)}^1$ for particle~$1$ in the chain \chainthree as a function of $\theta_{(23)1}$ and $\phi_{23}$.
Due to the low dimensionality of the problem, only the sum $\phi_{\chainthree(\chainone)}^1 + \psi_{\chainthree(\chainone)}^1$ is unambiguously defined; it is brought to the range $[-\pi, 3\pi]$.
Discontinuities of $2\pi$ are clearly visible.
This effect is a consequence of a definition of the kinematic variables $\tau_\chainthree$
used for describing the chain~$\chainthree$ amplitude.
All choices of $\tau$ are equivalent: at any given point of the phase space, $\tau_\chainthree$ can be mapped to $\tau_\chainone$.
However, a continuous domain of $\tau_\chainone$ is not mapped to a continuous domain of $\tau_\chainthree$ due to the definition of angles using four-vectors in Eq.~\eqref{eq:phi}, which are not sensitive to $2\pi$ phases.
The decay-chain amplitude $\mathcal{A}^\chainthree_{\lambda_0;\collective{\lambda'}}(\tau_\chainthree)$ might suddenly flip sign in the continuous domain of $\tau_\chainone$ while passing the discontinuity of the mapping.
The $2\pi$ discontinuity translates into a sign change in Eq.~\eqref{eq:Achain}, when the Wigner function is of non-integer-spin.
Since the full expression $\mathcal{A}^\chainthree_{\lambda_0;\collective{\lambda'}} W^\chainthree_{\collective{\lambda'};\collective{\lambda}}$ should be a continuous function of the angles,
the Wigner rotation has to compensate for the discontinuities in $\mathcal{A}^\chainthree_{\lambda_0;\collective{\lambda'}}$.
Figure~\ref{fig:helicity_jumps} demonstrates the correspondence:
wherever helicity angles go from $\pi \rightarrow -\pi$, the azimuthal part of Wigner rotations experiences a $2\pi$~jump.

While the azimuthal part of the Wigner rotation is cumbersome,
the polar part is straightforward and does not depend on the overall orientation of the system.
We have verified that the numerical values for the Wigner angles in Eq.~\eqref{eq:L.rot} match exactly the values from the analytic formulas of Ref.~\cite{JPAC:2019ufm}. When working with decays of multiplicity exceeding three, there are no simple analytic expressions for Wigner rotations.

\subsection{Particle ordering}
\label{ssec:particle-ordering}

A topology defines its ordering scheme \ie, how particle one and particle two are chosen.
Upon modifying the ordering scheme, three changes are in place:
opposite helicity angles, new Wigner rotation, and order of particles in the vertex spin-orbit coupling.
Let us start with the latter update.
The spin-orbit couplings, introduced in Eq.~\eqref{eq:ls},
have a convenient transformation property [see Eq.~(5.57) in Ref.~\cite{Martin:1970hmp}]:
\begin{align}
    \hnodegen{l_\nodeind s_\nodeind}{{\done\dtwo}} = (-1)^{l_\nodeind+s_\nodeind-j_\done-j_\dtwo}  \hnodegen{l_\nodeind s_\nodeind}{{\dtwo\done}}\,,
\end{align}
The helicity angles are affected as follows,
\begin{align*}
    \theta_{j} = \pi - \theta_{i}\,, \\
    \phi_{j}   = \phi_{i} \pm \pi\,,
\end{align*}
where the sign of the $\phi$ shift depends on the ranges of the azimuthal angle,
the physical $\phi$ always stays within $[-\pi,\pi]$, as also discussed in Ref.~\cite{Wang:2020giv}.
It impacts the reaction amplitude via the Wigner D-function in Eq.~\eqref{eq:Achain}.
When comparing a decay-chain amplitude for non-integer spin state with its order-swapped version, one finds a bizarre situation: due to the $2\pi$ difference of the $\phi$ shift, there is a relative sign factor that depends on $\phi$.
Such an unaccounted sign would show up in an interference between decay chains moving it from positive to negative values depending on the definition range of the azimuthal.
Fortunately, the Wigner rotations come with a complementary phase making things consistent again.
In fact, the $\phi$-dependent $2\pi$ shift appears in even numbers of Wigner rotations. This $2\pi$ shift does not translate into a phase for integer spin.
For both possible transitions with bosons $b$ and fermions $f$, $b\to ff$, and $f\to b f$,
a full cancellation of the $\phi$-dependent phase is achieved.

\section{Other formulations}\label{sec:other-formulations}

Alternative amplitude formulations known as \minusphi and the \textit{canonical} convention,
prescribe different ways of traversing the decay graph.
Both solid angles of decay products and Wigner rotations appear different in these approaches.
Nonetheless, once implemented correctly, these formulations are mathematically equivalent to each other.

The \minusphi convention was originally used by Jacob and Wick in their foundational paper~\cite{Jacob:1959at}.
It is adopted in many books and reviews, including~\cite{Martin:1970hmp,Richman:1984gh,Chung:1971ri}.
It adds an additional $R_z^{-1}(\phi_\nodeind)$ transformation to the rotation sequence in Eq.~\eqref{eq:hel}. Namely,
\begin{align}
    \nonumber
    \Lambda^{(-\phi)}_{\mother \leftarrow \done} & = R(\phi_{k},\theta_\nodeind, -\phi_\nodeind) \, B_z(\gamma_\nodeind^{\done})\,,          \\ \label{eq:minus_phi}
    \Lambda^{(-\phi)}_{\mother \leftarrow \dtwo} & = R(\phi_{k},\theta_\nodeind, -\phi_\nodeind) R_y(\pi)\,  B_z(\gamma_\nodeind^{\dtwo})\,.
\end{align}
The extra rotation impacts azimuthal angles in the daughter frames.
The full amplitude is computed using the same formulae as in the helicity formulation, Eqs.~\eqref{eq:Afull} and \eqref{eq:Achain},
given a new set of angles and the rotation function \mbox{$D^{j}_{\rho,\eta}(\phi, \theta, -\phi)$}.

The \textit{canonical} formulation deals with the spin projection to a fixed axis rather than the momentum direction.
\begin{align} \label{eq:canonical.state}
    \ket{p;j,m}_\text{can} = R B R^{-1} \ket{0;j,\lambda}
\end{align}
A pure boost is used to transit between frames, namely,
\begin{align*}
    \Lambda^{(\text{can.})}_{\mother \leftarrow \done} & =  R(\phi_\nodeind, \theta_\nodeind) B_z(\gamma_\nodeind^{\done}) R^{-1}(\phi_\nodeind, \theta_\nodeind)\,,                                 \\
    \Lambda^{(\text{can.})}_{\mother \leftarrow \dtwo} & =  R(\tilde{\phi}_\nodeind ,\tilde{\theta}_\nodeind) B_z(\gamma_\nodeind^{\dtwo}) R^{-1}(\tilde{\phi}_\nodeind, \tilde{\theta}_\nodeind)\,,
\end{align*}
where $(\tilde{\phi}_\nodeind, \tilde{\theta}_\nodeind)$ is the solid angle of particle $\dtwo$.

Using the cascade parametrization,
the full amplitude reads as a sum of the decay chains analogously to Eq.~\eqref{eq:Afull}.
\begin{align} \nonumber
    \mathcal{A}_{m_0;\collective{m}}(\tau_\reftext) & = \sum_{\collective{m'}} \sum_{\chainind} \mathcal{A}^\chainind_{m_0;\collective{m'}}(\tau_\chainind) \\ \label{eq:Afull.canonical}
                                                    & \qquad \times \WignerAll{m}(\tau_\chainind|\tau_\reftext) \,,
\end{align}
where $m_0$ is a spin projection of the decaying particle,
and $\collective{m}$ collectively denotes the spin projections of the final-state particles.
The canonical formulation of decay chain amplitudes differs from Eq.~\eqref{eq:Afull}.
The parametrization, as given by Refs.~\cite{Martin:1970hmp,Li:2022qff}, is
\begin{align*}
    \mathcal{A}^{\chainind}_{m_0;\collective{m}}(\tau_\chainind) & = \sum_{m_\text{int.}\dots} \prod_{k} Y_{{l_\nodeind}}^{m_{l_\nodeind}}(\varOmega_\nodeind^\chainind)  X_\nodeind (\mu_\nodeind ) \\
                                                                 & \quad \times \sum_{l_\nodeind s_\nodeind}
    \hnode{l_\nodeind  s_\nodeind}(\mu_{\mother},\mu_{\done},\mu_{\dtwo})                                                                                                                            \\                              & \quad \times
    \CG{j_{\done}}{m_{\done}}{j_{\dtwo}}{m_{\dtwo}}{s_\nodeind}{m_{s\nodeind}} \, \CG{l_\nodeind}{m_{lk}}{s_\nodeind}{m_{s\nodeind}}{j_{\mother}}{m_{\mother}}\,,
\end{align*}
where \mbox{$m_{s\nodeind} = m_{\done} + m_{\dtwo}$}, and \mbox{$m_{l\nodeind} = m_{\mother} - m_{s\nodeind}$}.
Angular dependence is expressed using spherical harmonics,
related to the Wigner D-functions as
\mbox{$Y_{l}^m(\varOmega) = \sqrt{(2l+1)/4\pi}\,D_{m,0}^{l*}(\phi,\theta ,0)$}.
The Wigner rotations enter the canonical formulations in Eq.~\eqref{eq:Afull.canonical} as well, despite the usage of pure boosts consistently throughout the amplitude construction.
The appearance of nontrivial Wigner rotations in this case is widely recognized as the Thomas precession
(see derivation on p.~215 of Ref.~\cite{Gourgoulhon:2013gua}).

\begin{figure}[htb]
    \centering
    \includegraphics[width=.4\textwidth]{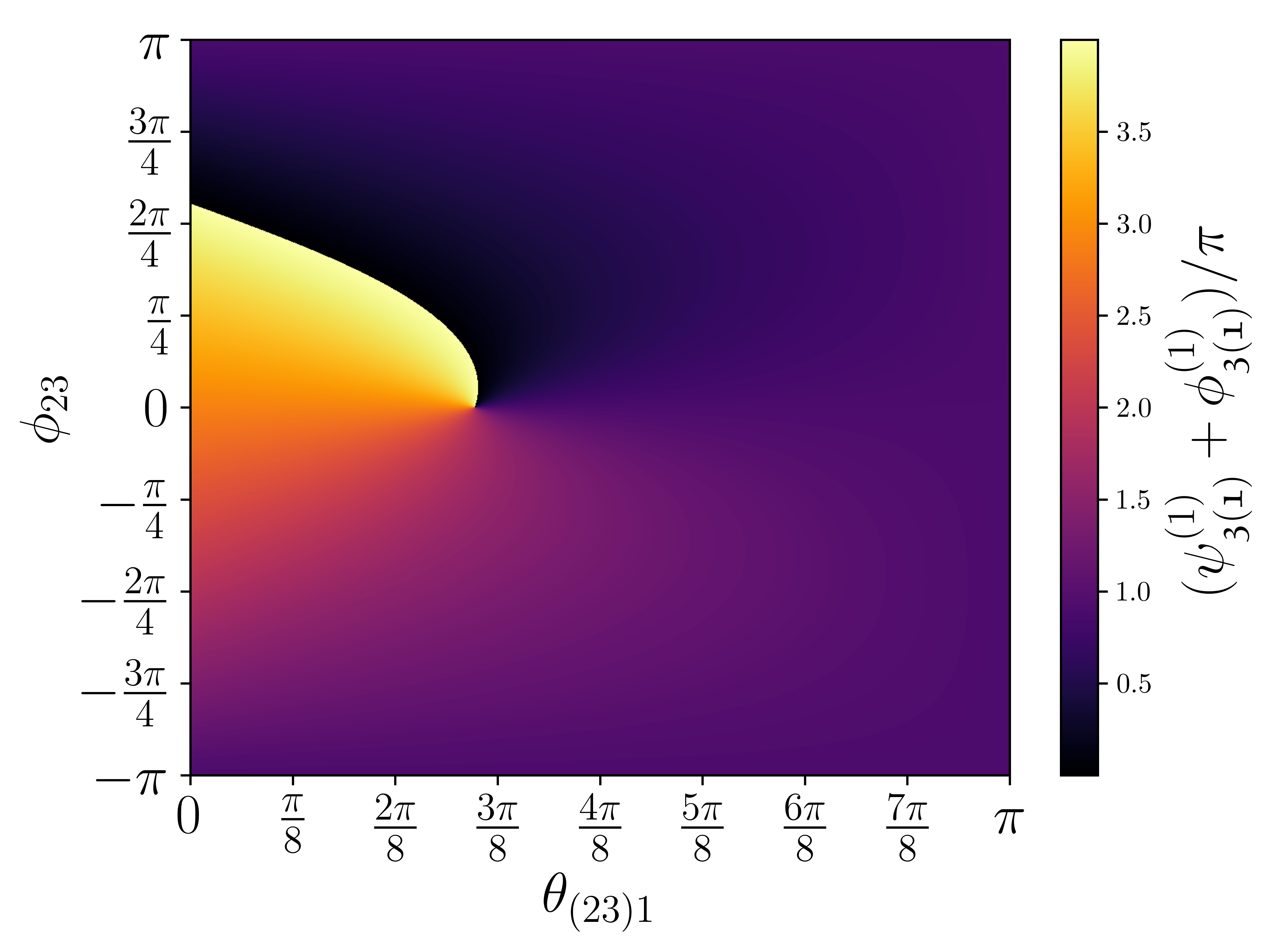}
    \caption{Azimuthal part of Wigner rotation, $\psi_{\chainthree(\chainone)}^1 + \phi_{\chainthree(\chainone)}^1$, for particle~$1$ in the chain $0 \to (12)3$, using the chain $0 \to (23)1$ as a reference computed in the \minusphi convention as a function of kinematic variables is depicted by the color. The $x$, and $y$ axes show the rotation angles $\theta_{(23)1}$ and $\phi_{(23)1}$, respectively, that are used to generate the four-vectors from the aligned configuration, $\phi_{(23)1}$ set to $1.4$.}
    \label{fig:minus_phi}
\end{figure}

\begin{figure}[htb]
    \centering
    \includegraphics[width=.4\textwidth]{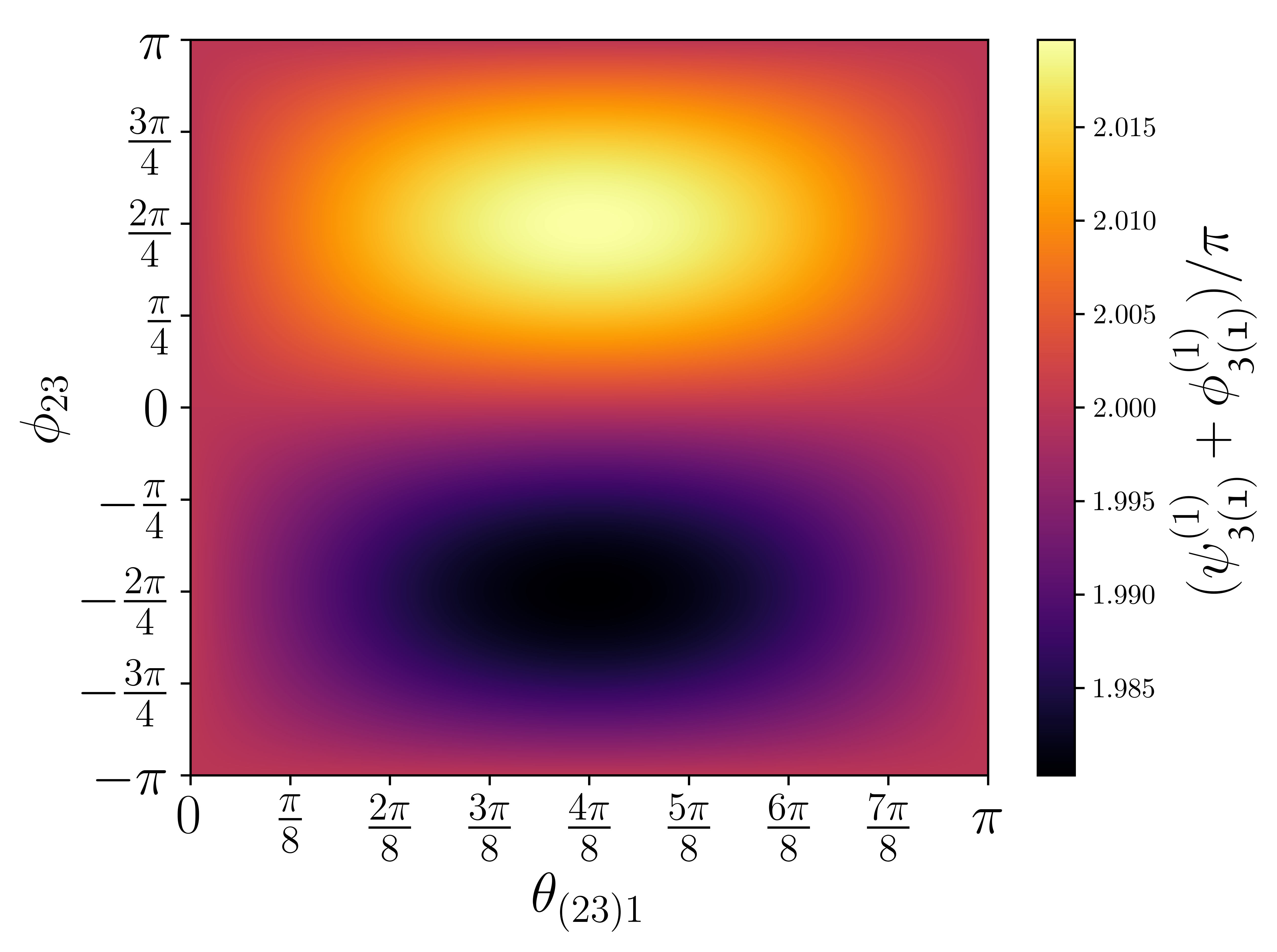}
    \caption{Azimuthal part of Wigner rotation, $\psi_{\chainthree(\chainone)}^1 + \phi_{\chainthree(\chainone)}^1$, for particle~$1$ in the chain $0 \to (12)3$, using the chain $0 \to (23)1$ as a reference computed in the \canonical convention as a function of kinematic variables, depicted by the color. The $x$ and $y$-axes show the rotation angles $\theta_{(23)1}$ and $\psi_{(23)1}$, respectively, that are used to generate the four-vectors from the aligned configuration, $\phi_{(23)1}$ set to $1.4$.}
    \label{fig:canonical}
\end{figure}

In Figure~\ref{fig:minus_phi}, we present the computation of the azimuthal Wigner rotation for chain 1 in the \minusphi convention, and in Figure~\ref{fig:canonical} the same for the \canonical convention is shown.
There appear no discontinuities of $2\pi$ in the azimuthal rotations for the \minusphi convention.

The visible structure is a jump by $4 \pi$ that has no effect on the value of the Wigner rotation.

Equivalence between the \helicity, \minusphi and \canonical formulations is far from being obvious.
There is a difference in $\phi$-dependent overall phase between the \helicity and \minusphi conventions, while it does not affect the differential distributions for unpolarized decay,
nor the interference patterns, as argued earlier.
The \canonical and \helicity amplitudes are expressed in different bases,
related by a linear transformation that accounts for the difference in Eqs.~\eqref{eq:hel.state} and \eqref{eq:canonical.state}.
We have obtained a numerical agreement between multiple physics cases as a part of the test set.
Particularly, a floating-point equivalence of the \helicity, \minusphi, and \canonical formulations
is demonstrated for unpolarized decays of a fermion, to a fermion, vector, and pseudoscalar ($1/2\to 1/2, 1, 0$) using a Python implementation~\texttt{decayangle}~\cite{habermann_2024_13741241} and a Julia package~\cite{mikhasenko_2024_13713312}.\\

The approach devised in this paper provides a general method for computing Wigner rotations in cascade decays. We established the applicability of our method across three distinct formulations, demonstrating their equivalence. Additionally, we provided the necessary tools and software to compute all angular variables required for amplitude construction, streamlining the analysis of hadronic decays. With this work, we aim to lower the barrier to entry for amplitude analysis and offer a comprehensive, concise framework for modeling cascade reactions.

\section*{Data availability}\label{sec:code-availability}

The numerical routine is available as a Python package for public use.
The code is public~\cite{habermann_2024_13741241} and installation via PyPI is possible.
The GitHub repository contains several notebooks exploring the different conventions and effects of rotations.

\bibliography{main}

\begin{thebibliography}{43}%
\makeatletter
\providecommand \@ifxundefined [1]{%
 \@ifx{#1\undefined}
}%
\providecommand \@ifnum [1]{%
 \ifnum #1\expandafter \@firstoftwo
 \else \expandafter \@secondoftwo
 \fi
}%
\providecommand \@ifx [1]{%
 \ifx #1\expandafter \@firstoftwo
 \else \expandafter \@secondoftwo
 \fi
}%
\providecommand \natexlab [1]{#1}%
\providecommand \enquote  [1]{``#1''}%
\providecommand \bibnamefont  [1]{#1}%
\providecommand \bibfnamefont [1]{#1}%
\providecommand \citenamefont [1]{#1}%
\providecommand \href@noop [0]{\@secondoftwo}%
\providecommand \href [0]{\begingroup \@sanitize@url \@href}%
\providecommand \@href[1]{\@@startlink{#1}\@@href}%
\providecommand \@@href[1]{\endgroup#1\@@endlink}%
\providecommand \@sanitize@url [0]{\catcode `\\12\catcode `\$12\catcode
  `\&12\catcode `\#12\catcode `\^12\catcode `\_12\catcode `\%12\relax}%
\providecommand \@@startlink[1]{}%
\providecommand \@@endlink[0]{}%
\providecommand \url  [0]{\begingroup\@sanitize@url \@url }%
\providecommand \@url [1]{\endgroup\@href {#1}{\urlprefix }}%
\providecommand \urlprefix  [0]{URL }%
\providecommand \Eprint [0]{\href }%
\providecommand \doibase [0]{https://doi.org/}%
\providecommand \selectlanguage [0]{\@gobble}%
\providecommand \bibinfo  [0]{\@secondoftwo}%
\providecommand \bibfield  [0]{\@secondoftwo}%
\providecommand \translation [1]{[#1]}%
\providecommand \BibitemOpen [0]{}%
\providecommand \bibitemStop [0]{}%
\providecommand \bibitemNoStop [0]{.\EOS\space}%
\providecommand \EOS [0]{\spacefactor3000\relax}%
\providecommand \BibitemShut  [1]{\csname bibitem#1\endcsname}%
\let\auto@bib@innerbib\@empty
\bibitem [{\citenamefont {Karliner}\ \emph {et~al.}(2018)\citenamefont
  {Karliner}, \citenamefont {Rosner},\ and\ \citenamefont
  {Skwarnicki}}]{Karliner:2017qhf}%
  \BibitemOpen
  \bibfield  {author} {\bibinfo {author} {\bibfnamefont {M.}~\bibnamefont
  {Karliner}}, \bibinfo {author} {\bibfnamefont {J.~L.}\ \bibnamefont
  {Rosner}},\ and\ \bibinfo {author} {\bibfnamefont {T.}~\bibnamefont
  {Skwarnicki}},\ }\bibfield  {title} {\bibinfo {title} {{Multiquark States}},\
  }\href {https://doi.org/10.1146/annurev-nucl-101917-020902} {\bibfield
  {journal} {\bibinfo  {journal} {Ann. Rev. Nucl. Part. Sci.}\ }\textbf
  {\bibinfo {volume} {68}},\ \bibinfo {pages} {17} (\bibinfo {year} {2018})},\
  \Eprint {https://arxiv.org/abs/1711.10626} {arXiv:1711.10626 [hep-ph]}
  \BibitemShut {NoStop}%
\bibitem [{\citenamefont {Klempt}\ and\ \citenamefont
  {Richard}(2010)}]{Klempt:2009pi}%
  \BibitemOpen
  \bibfield  {author} {\bibinfo {author} {\bibfnamefont {E.}~\bibnamefont
  {Klempt}}\ and\ \bibinfo {author} {\bibfnamefont {J.-M.}\ \bibnamefont
  {Richard}},\ }\bibfield  {title} {\bibinfo {title} {{Baryon spectroscopy}},\
  }\href {https://doi.org/10.1103/RevModPhys.82.1095} {\bibfield  {journal}
  {\bibinfo  {journal} {Rev. Mod. Phys.}\ }\textbf {\bibinfo {volume} {82}},\
  \bibinfo {pages} {1095} (\bibinfo {year} {2010})},\ \Eprint
  {https://arxiv.org/abs/0901.2055} {arXiv:0901.2055 [hep-ph]} \BibitemShut
  {NoStop}%
\bibitem [{\citenamefont {Richard}(2016)}]{Richard:2016eis}%
  \BibitemOpen
  \bibfield  {author} {\bibinfo {author} {\bibfnamefont {J.-M.}\ \bibnamefont
  {Richard}},\ }\bibfield  {title} {\bibinfo {title} {{Exotic hadrons: review
  and perspectives}},\ }\href {https://doi.org/10.1007/s00601-016-1159-0}
  {\bibfield  {journal} {\bibinfo  {journal} {Few Body Syst.}\ }\textbf
  {\bibinfo {volume} {57}},\ \bibinfo {pages} {1185} (\bibinfo {year}
  {2016})},\ \Eprint {https://arxiv.org/abs/1606.08593} {arXiv:1606.08593
  [hep-ph]} \BibitemShut {NoStop}%
\bibitem [{\citenamefont {Guo}\ \emph {et~al.}(2018)\citenamefont {Guo},
  \citenamefont {Hanhart}, \citenamefont {Mei\ss{}ner}, \citenamefont {Wang},
  \citenamefont {Zhao},\ and\ \citenamefont {Zou}}]{Guo:2017jvc}%
  \BibitemOpen
  \bibfield  {author} {\bibinfo {author} {\bibfnamefont {F.-K.}\ \bibnamefont
  {Guo}}, \bibinfo {author} {\bibfnamefont {C.}~\bibnamefont {Hanhart}},
  \bibinfo {author} {\bibfnamefont {U.-G.}\ \bibnamefont {Mei\ss{}ner}},
  \bibinfo {author} {\bibfnamefont {Q.}~\bibnamefont {Wang}}, \bibinfo {author}
  {\bibfnamefont {Q.}~\bibnamefont {Zhao}},\ and\ \bibinfo {author}
  {\bibfnamefont {B.-S.}\ \bibnamefont {Zou}},\ }\bibfield  {title} {\bibinfo
  {title} {{Hadronic molecules}},\ }\href
  {https://doi.org/10.1103/RevModPhys.90.015004} {\bibfield  {journal}
  {\bibinfo  {journal} {Rev. Mod. Phys.}\ }\textbf {\bibinfo {volume} {90}},\
  \bibinfo {pages} {015004} (\bibinfo {year} {2018})},\ \bibinfo {note}
  {[Erratum: Rev.Mod.Phys. 94, 029901 (2022)]},\ \Eprint
  {https://arxiv.org/abs/1705.00141} {arXiv:1705.00141 [hep-ph]} \BibitemShut
  {NoStop}%
\bibitem [{\citenamefont {Esposito}\ \emph {et~al.}(2017)\citenamefont
  {Esposito}, \citenamefont {Pilloni},\ and\ \citenamefont
  {Polosa}}]{Esposito:2016noz}%
  \BibitemOpen
  \bibfield  {author} {\bibinfo {author} {\bibfnamefont {A.}~\bibnamefont
  {Esposito}}, \bibinfo {author} {\bibfnamefont {A.}~\bibnamefont {Pilloni}},\
  and\ \bibinfo {author} {\bibfnamefont {A.~D.}\ \bibnamefont {Polosa}},\
  }\bibfield  {title} {\bibinfo {title} {{Multiquark Resonances}},\ }\href
  {https://doi.org/10.1016/j.physrep.2016.11.002} {\bibfield  {journal}
  {\bibinfo  {journal} {Phys. Rept.}\ }\textbf {\bibinfo {volume} {668}},\
  \bibinfo {pages} {1} (\bibinfo {year} {2017})},\ \Eprint
  {https://arxiv.org/abs/1611.07920} {arXiv:1611.07920 [hep-ph]} \BibitemShut
  {NoStop}%
\bibitem [{\citenamefont {Aaij}\ \emph {et~al.}(2015)\citenamefont {Aaij} \emph
  {et~al.}}]{LHCb:2015yax}%
  \BibitemOpen
  \bibfield  {author} {\bibinfo {author} {\bibfnamefont {R.}~\bibnamefont
  {Aaij}} \emph {et~al.} (\bibinfo {collaboration} {LHCb}),\ }\bibfield
  {title} {\bibinfo {title} {{Observation of $J/\psi p$ Resonances Consistent
  with Pentaquark States in $\Lambda_b^0 \to J/\psi K^- p$ Decays}},\ }\href
  {https://doi.org/10.1103/PhysRevLett.115.072001} {\bibfield  {journal}
  {\bibinfo  {journal} {Phys. Rev. Lett.}\ }\textbf {\bibinfo {volume} {115}},\
  \bibinfo {pages} {072001} (\bibinfo {year} {2015})},\ \Eprint
  {https://arxiv.org/abs/1507.03414} {arXiv:1507.03414 [hep-ex]} \BibitemShut
  {NoStop}%
\bibitem [{\citenamefont {Aaij}\ \emph {et~al.}(2019)\citenamefont {Aaij} \emph
  {et~al.}}]{LHCb:2019kea}%
  \BibitemOpen
  \bibfield  {author} {\bibinfo {author} {\bibfnamefont {R.}~\bibnamefont
  {Aaij}} \emph {et~al.} (\bibinfo {collaboration} {LHCb}),\ }\bibfield
  {title} {\bibinfo {title} {{Observation of a narrow pentaquark state,
  $P_c(4312)^+$, and of two-peak structure of the $P_c(4450)^+$}},\ }\href
  {https://doi.org/10.1103/PhysRevLett.122.222001} {\bibfield  {journal}
  {\bibinfo  {journal} {Phys. Rev. Lett.}\ }\textbf {\bibinfo {volume} {122}},\
  \bibinfo {pages} {222001} (\bibinfo {year} {2019})},\ \Eprint
  {https://arxiv.org/abs/1904.03947} {arXiv:1904.03947 [hep-ex]} \BibitemShut
  {NoStop}%
\bibitem [{\citenamefont {Aaij}\ \emph
  {et~al.}(2017{\natexlab{a}})\citenamefont {Aaij} \emph
  {et~al.}}]{LHCb:2016axx}%
  \BibitemOpen
  \bibfield  {author} {\bibinfo {author} {\bibfnamefont {R.}~\bibnamefont
  {Aaij}} \emph {et~al.} (\bibinfo {collaboration} {LHCb}),\ }\bibfield
  {title} {\bibinfo {title} {{Observation of $J/\psi\phi$ structures consistent
  with exotic states from amplitude analysis of $B^+\to J/\psi \phi K^+$
  decays}},\ }\href {https://doi.org/10.1103/PhysRevLett.118.022003} {\bibfield
   {journal} {\bibinfo  {journal} {Phys. Rev. Lett.}\ }\textbf {\bibinfo
  {volume} {118}},\ \bibinfo {pages} {022003} (\bibinfo {year}
  {2017}{\natexlab{a}})},\ \Eprint {https://arxiv.org/abs/1606.07895}
  {arXiv:1606.07895 [hep-ex]} \BibitemShut {NoStop}%
\bibitem [{\citenamefont {Aaij}\ \emph
  {et~al.}(2017{\natexlab{b}})\citenamefont {Aaij} \emph
  {et~al.}}]{LHCb:2016nsl}%
  \BibitemOpen
  \bibfield  {author} {\bibinfo {author} {\bibfnamefont {R.}~\bibnamefont
  {Aaij}} \emph {et~al.} (\bibinfo {collaboration} {LHCb}),\ }\bibfield
  {title} {\bibinfo {title} {{Amplitude analysis of $B^+\to J/\psi \phi K^+$
  decays}},\ }\href {https://doi.org/10.1103/PhysRevD.95.012002} {\bibfield
  {journal} {\bibinfo  {journal} {Phys. Rev. D}\ }\textbf {\bibinfo {volume}
  {95}},\ \bibinfo {pages} {012002} (\bibinfo {year} {2017}{\natexlab{b}})},\
  \Eprint {https://arxiv.org/abs/1606.07898} {arXiv:1606.07898 [hep-ex]}
  \BibitemShut {NoStop}%
\bibitem [{\citenamefont {Aaij}\ \emph {et~al.}(2021)\citenamefont {Aaij} \emph
  {et~al.}}]{LHCb:2021uow}%
  \BibitemOpen
  \bibfield  {author} {\bibinfo {author} {\bibfnamefont {R.}~\bibnamefont
  {Aaij}} \emph {et~al.} (\bibinfo {collaboration} {LHCb}),\ }\bibfield
  {title} {\bibinfo {title} {{Observation of New Resonances Decaying to $J/\psi
  K^+$+ and $J/\psi \phi$}},\ }\href
  {https://doi.org/10.1103/PhysRevLett.127.082001} {\bibfield  {journal}
  {\bibinfo  {journal} {Phys. Rev. Lett.}\ }\textbf {\bibinfo {volume} {127}},\
  \bibinfo {pages} {082001} (\bibinfo {year} {2021})},\ \Eprint
  {https://arxiv.org/abs/2103.01803} {arXiv:2103.01803 [hep-ex]} \BibitemShut
  {NoStop}%
\bibitem [{\citenamefont {Ablikim}\ \emph {et~al.}(2020)\citenamefont {Ablikim}
  \emph {et~al.}}]{BESIII:2020oph}%
  \BibitemOpen
  \bibfield  {author} {\bibinfo {author} {\bibfnamefont {M.}~\bibnamefont
  {Ablikim}} \emph {et~al.} (\bibinfo {collaboration} {BESIII}),\ }\bibfield
  {title} {\bibinfo {title} {{Study of the process $e^+e^-\to\pi^0\pi^0 J/\psi$
  and neutral charmonium-like state $Z_c(3900)^0$}},\ }\href
  {https://doi.org/10.1103/PhysRevD.102.012009} {\bibfield  {journal} {\bibinfo
   {journal} {Phys. Rev. D}\ }\textbf {\bibinfo {volume} {102}},\ \bibinfo
  {pages} {012009} (\bibinfo {year} {2020})},\ \Eprint
  {https://arxiv.org/abs/2004.13788} {arXiv:2004.13788 [hep-ex]} \BibitemShut
  {NoStop}%
\bibitem [{\citenamefont {Chilikin}\ \emph {et~al.}(2013)\citenamefont
  {Chilikin} \emph {et~al.}}]{PhysRevD.88.074026}%
  \BibitemOpen
  \bibfield  {author} {\bibinfo {author} {\bibfnamefont {K.}~\bibnamefont
  {Chilikin}} \emph {et~al.} (\bibinfo {collaboration} {Belle Collaboration}),\
  }\bibfield  {title} {\bibinfo {title} {Experimental constraints on the spin
  and parity of the $z\mathbf{(}4430{\mathbf{)}}^{\mathbf{+}}$},\ }\href
  {https://doi.org/10.1103/PhysRevD.88.074026} {\bibfield  {journal} {\bibinfo
  {journal} {Phys. Rev. D}\ }\textbf {\bibinfo {volume} {88}},\ \bibinfo
  {pages} {074026} (\bibinfo {year} {2013})}\BibitemShut {NoStop}%
\bibitem [{\citenamefont {Herndon}\ \emph {et~al.}(1975)\citenamefont
  {Herndon}, \citenamefont {Soding},\ and\ \citenamefont
  {Cashmore}}]{Herndon:1973yn}%
  \BibitemOpen
  \bibfield  {author} {\bibinfo {author} {\bibfnamefont {D.}~\bibnamefont
  {Herndon}}, \bibinfo {author} {\bibfnamefont {P.}~\bibnamefont {Soding}},\
  and\ \bibinfo {author} {\bibfnamefont {R.~J.}\ \bibnamefont {Cashmore}},\
  }\bibfield  {title} {\bibinfo {title} {{A generalized Isobar Model
  Formalism}},\ }\href {https://doi.org/10.1103/PhysRevD.11.3165} {\bibfield
  {journal} {\bibinfo  {journal} {Phys. Rev. D}\ }\textbf {\bibinfo {volume}
  {11}},\ \bibinfo {pages} {3165} (\bibinfo {year} {1975})}\BibitemShut
  {NoStop}%
\bibitem [{\citenamefont {Hansen}\ \emph {et~al.}(1974)\citenamefont {Hansen},
  \citenamefont {Jones}, \citenamefont {Otter},\ and\ \citenamefont
  {Rudolph}}]{Hansen:1973gb}%
  \BibitemOpen
  \bibfield  {author} {\bibinfo {author} {\bibfnamefont {J.}~\bibnamefont
  {Hansen}}, \bibinfo {author} {\bibfnamefont {G.}~\bibnamefont {Jones}},
  \bibinfo {author} {\bibfnamefont {G.}~\bibnamefont {Otter}},\ and\ \bibinfo
  {author} {\bibfnamefont {G.}~\bibnamefont {Rudolph}},\ }\bibfield  {title}
  {\bibinfo {title} {Formalism and assumptions involved in partial-wave
  analysis of three-meson systems},\ }\href
  {https://doi.org/https://doi.org/10.1016/0550-3213(74)90241-7} {\bibfield
  {journal} {\bibinfo  {journal} {Nuclear Physics B}\ }\textbf {\bibinfo
  {volume} {81}},\ \bibinfo {pages} {403} (\bibinfo {year} {1974})}\BibitemShut
  {NoStop}%
\bibitem [{\citenamefont {Aitchison}(1972)}]{Aitchison:1972ay}%
  \BibitemOpen
  \bibfield  {author} {\bibinfo {author} {\bibfnamefont {I.~J.~R.}\
  \bibnamefont {Aitchison}},\ }\bibfield  {title} {\bibinfo {title} {{K-Matrix
  Formalism for overlapping Resonances}},\ }\href
  {https://doi.org/10.1016/0375-9474(72)90305-3} {\bibfield  {journal}
  {\bibinfo  {journal} {Nucl. Phys. A}\ }\textbf {\bibinfo {volume} {189}},\
  \bibinfo {pages} {417} (\bibinfo {year} {1972})}\BibitemShut {NoStop}%
\bibitem [{\citenamefont {Workman}\ \emph {et~al.}(2022)\citenamefont {Workman}
  \emph {et~al.}}]{ParticleDataGroup:2022pth}%
  \BibitemOpen
  \bibfield  {author} {\bibinfo {author} {\bibfnamefont {R.~L.}\ \bibnamefont
  {Workman}} \emph {et~al.} (\bibinfo {collaboration} {Particle Data Group}),\
  }\bibfield  {title} {\bibinfo {title} {{Review of Particle Physics}},\ }\href
  {https://doi.org/10.1093/ptep/ptac097} {\bibfield  {journal} {\bibinfo
  {journal} {PTEP}\ }\textbf {\bibinfo {volume} {2022}},\ \bibinfo {pages}
  {083C01} (\bibinfo {year} {2022})}\BibitemShut {NoStop}%
\bibitem [{\citenamefont {Link}\ \emph {et~al.}(2004)\citenamefont {Link} \emph
  {et~al.}}]{FOCUS:2003tdy}%
  \BibitemOpen
  \bibfield  {author} {\bibinfo {author} {\bibfnamefont {J.~M.}\ \bibnamefont
  {Link}} \emph {et~al.} (\bibinfo {collaboration} {FOCUS}),\ }\bibfield
  {title} {\bibinfo {title} {{Dalitz plot analysis of D(s)+ and D+ decay to pi+
  pi- pi+ using the K matrix formalism}},\ }\href
  {https://doi.org/10.1016/j.physletb.2004.01.065} {\bibfield  {journal}
  {\bibinfo  {journal} {Phys. Lett. B}\ }\textbf {\bibinfo {volume} {585}},\
  \bibinfo {pages} {200} (\bibinfo {year} {2004})},\ \Eprint
  {https://arxiv.org/abs/hep-ex/0312040} {arXiv:hep-ex/0312040} \BibitemShut
  {NoStop}%
\bibitem [{\citenamefont {Aaij}\ \emph
  {et~al.}(2024{\natexlab{a}})\citenamefont {Aaij} \emph
  {et~al.}}]{LHCb:2024cwp}%
  \BibitemOpen
  \bibfield  {author} {\bibinfo {author} {\bibfnamefont {R.}~\bibnamefont
  {Aaij}} \emph {et~al.} (\bibinfo {collaboration} {LHCb}),\ }\bibfield
  {title} {\bibinfo {title} {{Amplitude analysis of $B^+ \to \psi(2S) K^+ \pi^+
  \pi^-$ decays}},\ }\href@noop {} {\  (\bibinfo {year}
  {2024}{\natexlab{a}})},\ \Eprint {https://arxiv.org/abs/2407.12475}
  {arXiv:2407.12475 [hep-ex]} \BibitemShut {NoStop}%
\bibitem [{\citenamefont {Khuri}\ and\ \citenamefont
  {Treiman}(1960)}]{Khuri:1960zz}%
  \BibitemOpen
  \bibfield  {author} {\bibinfo {author} {\bibfnamefont {N.~N.}\ \bibnamefont
  {Khuri}}\ and\ \bibinfo {author} {\bibfnamefont {S.~B.}\ \bibnamefont
  {Treiman}},\ }\bibfield  {title} {\bibinfo {title} {{Pion-Pion Scattering and
  $K^{\pm}\to 3\pi$ Decay}},\ }\href {https://doi.org/10.1103/PhysRev.119.1115}
  {\bibfield  {journal} {\bibinfo  {journal} {Phys. Rev.}\ }\textbf {\bibinfo
  {volume} {119}},\ \bibinfo {pages} {1115} (\bibinfo {year}
  {1960})}\BibitemShut {NoStop}%
\bibitem [{\citenamefont {Aitchison}\ and\ \citenamefont
  {Brehm}(1979)}]{Aitchison:1979fj}%
  \BibitemOpen
  \bibfield  {author} {\bibinfo {author} {\bibfnamefont {I.~J.~R.}\
  \bibnamefont {Aitchison}}\ and\ \bibinfo {author} {\bibfnamefont {J.~J.}\
  \bibnamefont {Brehm}},\ }\bibfield  {title} {\bibinfo {title} {{Are there
  important unitarity corrections to the isobar model?}},\ }\href
  {https://doi.org/10.1016/0370-2693(79)90056-X} {\bibfield  {journal}
  {\bibinfo  {journal} {Phys. Lett. B}\ }\textbf {\bibinfo {volume} {84}},\
  \bibinfo {pages} {349} (\bibinfo {year} {1979})}\BibitemShut {NoStop}%
\bibitem [{\citenamefont {Pasquier}\ and\ \citenamefont
  {Pasquier}(1968)}]{Pasquier:1968zz}%
  \BibitemOpen
  \bibfield  {author} {\bibinfo {author} {\bibfnamefont {R.}~\bibnamefont
  {Pasquier}}\ and\ \bibinfo {author} {\bibfnamefont {J.~Y.}\ \bibnamefont
  {Pasquier}},\ }\bibfield  {title} {\bibinfo {title} {{Khuri-Treiman-Type
  Equations for Three-Body Decay and Production Processes}},\ }\href
  {https://doi.org/10.1103/PhysRev.170.1294} {\bibfield  {journal} {\bibinfo
  {journal} {Phys. Rev.}\ }\textbf {\bibinfo {volume} {170}},\ \bibinfo {pages}
  {1294} (\bibinfo {year} {1968})}\BibitemShut {NoStop}%
\bibitem [{\citenamefont {Altmannshofer}\ \emph {et~al.}(2009)\citenamefont
  {Altmannshofer}, \citenamefont {Ball}, \citenamefont {Bharucha},
  \citenamefont {Buras}, \citenamefont {Straub},\ and\ \citenamefont
  {Wick}}]{Altmannshofer:2008dz}%
  \BibitemOpen
  \bibfield  {author} {\bibinfo {author} {\bibfnamefont {W.}~\bibnamefont
  {Altmannshofer}}, \bibinfo {author} {\bibfnamefont {P.}~\bibnamefont {Ball}},
  \bibinfo {author} {\bibfnamefont {A.}~\bibnamefont {Bharucha}}, \bibinfo
  {author} {\bibfnamefont {A.~J.}\ \bibnamefont {Buras}}, \bibinfo {author}
  {\bibfnamefont {D.~M.}\ \bibnamefont {Straub}},\ and\ \bibinfo {author}
  {\bibfnamefont {M.}~\bibnamefont {Wick}},\ }\bibfield  {title} {\bibinfo
  {title} {{Symmetries and Asymmetries of $B \to K^{*} \mu^{+} \mu^{-}$ Decays
  in the Standard Model and Beyond}},\ }\href
  {https://doi.org/10.1088/1126-6708/2009/01/019} {\bibfield  {journal}
  {\bibinfo  {journal} {JHEP}\ }\textbf {\bibinfo {volume} {01}},\ \bibinfo
  {pages} {019}},\ \Eprint {https://arxiv.org/abs/0811.1214} {arXiv:0811.1214
  [hep-ph]} \BibitemShut {NoStop}%
\bibitem [{\citenamefont {Aaij}\ \emph
  {et~al.}(2024{\natexlab{b}})\citenamefont {Aaij} \emph
  {et~al.}}]{LHCb:2024onj}%
  \BibitemOpen
  \bibfield  {author} {\bibinfo {author} {\bibfnamefont {R.}~\bibnamefont
  {Aaij}} \emph {et~al.} (\bibinfo {collaboration} {LHCb}),\ }\bibfield
  {title} {\bibinfo {title} {{Comprehensive analysis of local and nonlocal
  amplitudes in the $B^0\rightarrow K^{*0}\mu^+\mu^-$ decay}},\ }\href@noop {}
  {\  (\bibinfo {year} {2024}{\natexlab{b}})},\ \Eprint
  {https://arxiv.org/abs/2405.17347} {arXiv:2405.17347 [hep-ex]} \BibitemShut
  {NoStop}%
\bibitem [{\citenamefont {Filippini}\ \emph {et~al.}(1995)\citenamefont
  {Filippini}, \citenamefont {Fontana},\ and\ \citenamefont
  {Rotondi}}]{Filippini:1995yc}%
  \BibitemOpen
  \bibfield  {author} {\bibinfo {author} {\bibfnamefont {V.}~\bibnamefont
  {Filippini}}, \bibinfo {author} {\bibfnamefont {A.}~\bibnamefont {Fontana}},\
  and\ \bibinfo {author} {\bibfnamefont {A.}~\bibnamefont {Rotondi}},\
  }\bibfield  {title} {\bibinfo {title} {{Covariant spin tensors in meson
  spectroscopy}},\ }\href {https://doi.org/10.1103/PhysRevD.51.2247} {\bibfield
   {journal} {\bibinfo  {journal} {Phys. Rev. D}\ }\textbf {\bibinfo {volume}
  {51}},\ \bibinfo {pages} {2247} (\bibinfo {year} {1995})}\BibitemShut
  {NoStop}%
\bibitem [{\citenamefont {Chung}\ and\ \citenamefont
  {Friedrich}(2008)}]{Chung:2007nn}%
  \BibitemOpen
  \bibfield  {author} {\bibinfo {author} {\bibfnamefont {S.-U.}\ \bibnamefont
  {Chung}}\ and\ \bibinfo {author} {\bibfnamefont {J.}~\bibnamefont
  {Friedrich}},\ }\bibfield  {title} {\bibinfo {title} {{Covariant
  helicity-coupling amplitudes: A New formulation}},\ }\href
  {https://doi.org/10.1103/PhysRevD.78.074027} {\bibfield  {journal} {\bibinfo
  {journal} {Phys. Rev. D}\ }\textbf {\bibinfo {volume} {78}},\ \bibinfo
  {pages} {074027} (\bibinfo {year} {2008})},\ \Eprint
  {https://arxiv.org/abs/0711.3143} {arXiv:0711.3143 [hep-ph]} \BibitemShut
  {NoStop}%
\bibitem [{\citenamefont {Mikhasenko}\ \emph {et~al.}(2018)\citenamefont
  {Mikhasenko}, \citenamefont {Pilloni}, \citenamefont {Nys}, \citenamefont
  {Albaladejo}, \citenamefont {Fernandez-Ramirez}, \citenamefont {Jackura},
  \citenamefont {Mathieu}, \citenamefont {Sherrill}, \citenamefont
  {Skwarnicki},\ and\ \citenamefont {Szczepaniak}}]{JPAC:2017vtd}%
  \BibitemOpen
  \bibfield  {author} {\bibinfo {author} {\bibfnamefont {M.}~\bibnamefont
  {Mikhasenko}}, \bibinfo {author} {\bibfnamefont {A.}~\bibnamefont {Pilloni}},
  \bibinfo {author} {\bibfnamefont {J.}~\bibnamefont {Nys}}, \bibinfo {author}
  {\bibfnamefont {M.}~\bibnamefont {Albaladejo}}, \bibinfo {author}
  {\bibfnamefont {C.}~\bibnamefont {Fernandez-Ramirez}}, \bibinfo {author}
  {\bibfnamefont {A.}~\bibnamefont {Jackura}}, \bibinfo {author} {\bibfnamefont
  {V.}~\bibnamefont {Mathieu}}, \bibinfo {author} {\bibfnamefont
  {N.}~\bibnamefont {Sherrill}}, \bibinfo {author} {\bibfnamefont
  {T.}~\bibnamefont {Skwarnicki}},\ and\ \bibinfo {author} {\bibfnamefont
  {A.~P.}\ \bibnamefont {Szczepaniak}} (\bibinfo {collaboration} {JPAC}),\
  }\bibfield  {title} {\bibinfo {title} {{What is the right formalism to search
  for resonances?}},\ }\href {https://doi.org/10.1140/epjc/s10052-018-5670-y}
  {\bibfield  {journal} {\bibinfo  {journal} {Eur. Phys. J. C}\ }\textbf
  {\bibinfo {volume} {78}},\ \bibinfo {pages} {229} (\bibinfo {year} {2018})},\
  \Eprint {https://arxiv.org/abs/1712.02815} {arXiv:1712.02815 [hep-ph]}
  \BibitemShut {NoStop}%
\bibitem [{\citenamefont {Pilloni}\ \emph {et~al.}(2018)\citenamefont
  {Pilloni}, \citenamefont {Nys}, \citenamefont {Mikhasenko}, \citenamefont
  {Albaladejo}, \citenamefont {Fern{\'a}ndez-Ram{\'i}rez}, \citenamefont
  {Jackura}, \citenamefont {Mathieu}, \citenamefont {Sherrill}, \citenamefont
  {Skwarnicki},\ and\ \citenamefont {Szczepaniak}}]{JPAC:2018dfc}%
  \BibitemOpen
  \bibfield  {author} {\bibinfo {author} {\bibfnamefont {A.}~\bibnamefont
  {Pilloni}}, \bibinfo {author} {\bibfnamefont {J.}~\bibnamefont {Nys}},
  \bibinfo {author} {\bibfnamefont {M.}~\bibnamefont {Mikhasenko}}, \bibinfo
  {author} {\bibfnamefont {M.}~\bibnamefont {Albaladejo}}, \bibinfo {author}
  {\bibfnamefont {C.}~\bibnamefont {Fern{\'a}ndez-Ram{\'i}rez}}, \bibinfo
  {author} {\bibfnamefont {A.}~\bibnamefont {Jackura}}, \bibinfo {author}
  {\bibfnamefont {V.}~\bibnamefont {Mathieu}}, \bibinfo {author} {\bibfnamefont
  {N.}~\bibnamefont {Sherrill}}, \bibinfo {author} {\bibfnamefont
  {T.}~\bibnamefont {Skwarnicki}},\ and\ \bibinfo {author} {\bibfnamefont
  {A.~P.}\ \bibnamefont {Szczepaniak}} (\bibinfo {collaboration} {JPAC}),\
  }\bibfield  {title} {\bibinfo {title} {{What is the right formalism to search
  for resonances? II. The pentaquark chain}},\ }\href
  {https://doi.org/10.1140/epjc/s10052-018-6177-2} {\bibfield  {journal}
  {\bibinfo  {journal} {Eur. Phys. J. C}\ }\textbf {\bibinfo {volume} {78}},\
  \bibinfo {pages} {727} (\bibinfo {year} {2018})},\ \Eprint
  {https://arxiv.org/abs/1805.02113} {arXiv:1805.02113 [hep-ph]} \BibitemShut
  {NoStop}%
\bibitem [{\citenamefont {Chen}\ and\ \citenamefont
  {Ping}(2017)}]{Chen:2017gtx}%
  \BibitemOpen
  \bibfield  {author} {\bibinfo {author} {\bibfnamefont {H.}~\bibnamefont
  {Chen}}\ and\ \bibinfo {author} {\bibfnamefont {R.-G.}\ \bibnamefont
  {Ping}},\ }\bibfield  {title} {\bibinfo {title} {{Coherent helicity amplitude
  for sequential decays}},\ }\href {https://doi.org/10.1103/PhysRevD.95.076010}
  {\bibfield  {journal} {\bibinfo  {journal} {Phys. Rev. D}\ }\textbf {\bibinfo
  {volume} {95}},\ \bibinfo {pages} {076010} (\bibinfo {year} {2017})},\
  \Eprint {https://arxiv.org/abs/1704.05184} {arXiv:1704.05184 [hep-ph]}
  \BibitemShut {NoStop}%
\bibitem [{\citenamefont {Li}\ \emph {et~al.}(2023)\citenamefont {Li},
  \citenamefont {Dong},\ and\ \citenamefont {Jing}}]{Li:2022qff}%
  \BibitemOpen
  \bibfield  {author} {\bibinfo {author} {\bibfnamefont {X.-y.}\ \bibnamefont
  {Li}}, \bibinfo {author} {\bibfnamefont {X.-K.}\ \bibnamefont {Dong}},\ and\
  \bibinfo {author} {\bibfnamefont {H.-J.}\ \bibnamefont {Jing}},\ }\bibfield
  {title} {\bibinfo {title} {{Spin-orbit amplitudes for decays with arbitrary
  spin}},\ }\href {https://doi.org/10.1016/j.nuclphysa.2023.122761} {\bibfield
  {journal} {\bibinfo  {journal} {Nucl. Phys. A}\ }\textbf {\bibinfo {volume}
  {1040}},\ \bibinfo {pages} {122761} (\bibinfo {year} {2023})},\ \Eprint
  {https://arxiv.org/abs/2212.06417} {arXiv:2212.06417 [hep-ph]} \BibitemShut
  {NoStop}%
\bibitem [{\citenamefont {Marangotto}(2020)}]{Marangotto:2019ucc}%
  \BibitemOpen
  \bibfield  {author} {\bibinfo {author} {\bibfnamefont {D.}~\bibnamefont
  {Marangotto}},\ }\bibfield  {title} {\bibinfo {title} {{Helicity Amplitudes
  for Generic Multibody Particle Decays Featuring Multiple Decay Chains}},\
  }\href {https://doi.org/10.1155/2020/6674595} {\bibfield  {journal} {\bibinfo
   {journal} {Adv. High Energy Phys.}\ }\textbf {\bibinfo {volume} {2020}},\
  \bibinfo {pages} {6674595} (\bibinfo {year} {2020})},\ \Eprint
  {https://arxiv.org/abs/1911.10025} {arXiv:1911.10025 [hep-ph]} \BibitemShut
  {NoStop}%
\bibitem [{\citenamefont {Mikhasenko}\ \emph {et~al.}(2020)\citenamefont
  {Mikhasenko} \emph {et~al.}}]{JPAC:2019ufm}%
  \BibitemOpen
  \bibfield  {author} {\bibinfo {author} {\bibfnamefont {M.}~\bibnamefont
  {Mikhasenko}} \emph {et~al.} (\bibinfo {collaboration} {JPAC}),\ }\bibfield
  {title} {\bibinfo {title} {{Dalitz-plot decomposition for three-body
  decays}},\ }\href {https://doi.org/10.1103/PhysRevD.101.034033} {\bibfield
  {journal} {\bibinfo  {journal} {Phys. Rev. D}\ }\textbf {\bibinfo {volume}
  {101}},\ \bibinfo {pages} {034033} (\bibinfo {year} {2020})},\ \Eprint
  {https://arxiv.org/abs/1910.04566} {arXiv:1910.04566 [hep-ph]} \BibitemShut
  {NoStop}%
\bibitem [{\citenamefont {Wang}\ \emph {et~al.}(2021)\citenamefont {Wang},
  \citenamefont {Jiang}, \citenamefont {Liu}, \citenamefont {Qian},
  \citenamefont {Lyu},\ and\ \citenamefont {Zhang}}]{Wang:2020giv}%
  \BibitemOpen
  \bibfield  {author} {\bibinfo {author} {\bibfnamefont {M.}~\bibnamefont
  {Wang}}, \bibinfo {author} {\bibfnamefont {Y.}~\bibnamefont {Jiang}},
  \bibinfo {author} {\bibfnamefont {Y.}~\bibnamefont {Liu}}, \bibinfo {author}
  {\bibfnamefont {W.}~\bibnamefont {Qian}}, \bibinfo {author} {\bibfnamefont
  {X.}~\bibnamefont {Lyu}},\ and\ \bibinfo {author} {\bibfnamefont
  {L.}~\bibnamefont {Zhang}},\ }\bibfield  {title} {\bibinfo {title} {{A novel
  method to test particle ordering and final state alignment in helicity
  formalism}},\ }\href {https://doi.org/10.1088/1674-1137/abf139} {\bibfield
  {journal} {\bibinfo  {journal} {Chin. Phys. C}\ }\textbf {\bibinfo {volume}
  {45}},\ \bibinfo {pages} {063103} (\bibinfo {year} {2021})},\ \Eprint
  {https://arxiv.org/abs/2012.03699} {arXiv:2012.03699 [hep-ex]} \BibitemShut
  {NoStop}%
\bibitem [{\citenamefont {Anisovich}\ and\ \citenamefont
  {Sarantsev}(2006)}]{Anisovich:2006bc}%
  \BibitemOpen
  \bibfield  {author} {\bibinfo {author} {\bibfnamefont {A.~V.}\ \bibnamefont
  {Anisovich}}\ and\ \bibinfo {author} {\bibfnamefont {A.~V.}\ \bibnamefont
  {Sarantsev}},\ }\bibfield  {title} {\bibinfo {title} {{Partial decay widths
  of baryons in the spin-momentum operator expansion method}},\ }\href
  {https://doi.org/10.1140/epja/i2006-10102-1} {\bibfield  {journal} {\bibinfo
  {journal} {Eur. Phys. J. A}\ }\textbf {\bibinfo {volume} {30}},\ \bibinfo
  {pages} {427} (\bibinfo {year} {2006})},\ \Eprint
  {https://arxiv.org/abs/hep-ph/0605135} {arXiv:hep-ph/0605135} \BibitemShut
  {NoStop}%
\bibitem [{\citenamefont {Zemach}(1965)}]{Zemach:1965ycj}%
  \BibitemOpen
  \bibfield  {author} {\bibinfo {author} {\bibfnamefont {C.}~\bibnamefont
  {Zemach}},\ }\bibfield  {title} {\bibinfo {title} {{Use of angular momentum
  tensors}},\ }\href {https://doi.org/10.1103/PhysRev.140.B97} {\bibfield
  {journal} {\bibinfo  {journal} {Phys. Rev.}\ }\textbf {\bibinfo {volume}
  {140}},\ \bibinfo {pages} {B97} (\bibinfo {year} {1965})}\BibitemShut
  {NoStop}%
\bibitem [{\citenamefont {Chung}(1998)}]{Chung:1997jn}%
  \BibitemOpen
  \bibfield  {author} {\bibinfo {author} {\bibfnamefont {S.~U.}\ \bibnamefont
  {Chung}},\ }\bibfield  {title} {\bibinfo {title} {{A General formulation of
  covariant helicity coupling amplitudes}},\ }\href
  {https://doi.org/10.1103/PhysRevD.57.431} {\bibfield  {journal} {\bibinfo
  {journal} {Phys. Rev. D}\ }\textbf {\bibinfo {volume} {57}},\ \bibinfo
  {pages} {431} (\bibinfo {year} {1998})}\BibitemShut {NoStop}%
\bibitem [{\citenamefont {Byckling}\ and\ \citenamefont
  {Kajantie}(1971)}]{Byckling:1971vca}%
  \BibitemOpen
  \bibfield  {author} {\bibinfo {author} {\bibfnamefont {E.}~\bibnamefont
  {Byckling}}\ and\ \bibinfo {author} {\bibfnamefont {K.}~\bibnamefont
  {Kajantie}},\ }\href@noop {} {\emph {\bibinfo {title} {{Particle Kinematics}:
  {(Chapters I-VI, X)}}}}\ (\bibinfo  {publisher} {University of Jyvaskyla},\
  \bibinfo {address} {Jyvaskyla, Finland},\ \bibinfo {year} {1971})\BibitemShut
  {NoStop}%
\bibitem [{\citenamefont {Martin}\ and\ \citenamefont
  {Spearman}(1970)}]{Martin:1970hmp}%
  \BibitemOpen
  \bibfield  {author} {\bibinfo {author} {\bibfnamefont {A.~D.}\ \bibnamefont
  {Martin}}\ and\ \bibinfo {author} {\bibfnamefont {T.~D.}\ \bibnamefont
  {Spearman}},\ }\href@noop {} {\emph {\bibinfo {title} {{Elementary Particle
  Theory}}}}\ (\bibinfo  {publisher} {North-Holland Publishing Co.},\ \bibinfo
  {address} {Amsterdam},\ \bibinfo {year} {1970})\BibitemShut {NoStop}%
\bibitem [{\citenamefont {Richman}(1984)}]{Richman:1984gh}%
  \BibitemOpen
  \bibfield  {author} {\bibinfo {author} {\bibfnamefont {J.~D.}\ \bibnamefont
  {Richman}},\ }\bibfield  {title} {\bibinfo {title} {{An Experimenter's Guide
  to the Helicity Formalism}},\ }\href@noop {} {\  (\bibinfo {year}
  {1984})}\BibitemShut {NoStop}%
\bibitem [{\citenamefont {Jacob}\ and\ \citenamefont
  {Wick}(1959)}]{Jacob:1959at}%
  \BibitemOpen
  \bibfield  {author} {\bibinfo {author} {\bibfnamefont {M.}~\bibnamefont
  {Jacob}}\ and\ \bibinfo {author} {\bibfnamefont {G.~C.}\ \bibnamefont
  {Wick}},\ }\bibfield  {title} {\bibinfo {title} {{On the General Theory of
  Collisions for Particles with Spin}},\ }\href
  {https://doi.org/10.1006/aphy.2000.6022} {\bibfield  {journal} {\bibinfo
  {journal} {Annals Phys.}\ }\textbf {\bibinfo {volume} {7}},\ \bibinfo {pages}
  {404} (\bibinfo {year} {1959})}\BibitemShut {NoStop}%
\bibitem [{\citenamefont {Chung}(1971)}]{Chung:1971ri}%
  \BibitemOpen
  \bibfield  {author} {\bibinfo {author} {\bibfnamefont {S.~U.}\ \bibnamefont
  {Chung}},\ }\bibfield  {title} {\bibinfo {title} {{Spin Formalisms}}\ }\href
  {https://doi.org/10.5170/CERN-1971-008} {10.5170/CERN-1971-008} (\bibinfo
  {year} {1971})\BibitemShut {NoStop}%
\bibitem [{\citenamefont {Gourgoulhon}(2013)}]{Gourgoulhon:2013gua}%
  \BibitemOpen
  \bibfield  {author} {\bibinfo {author} {\bibfnamefont {E.}~\bibnamefont
  {Gourgoulhon}},\ }\href {https://doi.org/10.1007/978-3-642-37276-6} {\emph
  {\bibinfo {title} {{Special Relativity in General Frames}}}},\ Graduate Texts
  in Physics\ (\bibinfo  {publisher} {Springer},\ \bibinfo {address} {Berlin,
  Heidelberg},\ \bibinfo {year} {2013})\BibitemShut {NoStop}%
\bibitem [{\citenamefont {Habermann}\ and\ \citenamefont
  {Mikhasenko}(2024)}]{habermann_2024_13741241}%
  \BibitemOpen
  \bibfield  {author} {\bibinfo {author} {\bibfnamefont {K.}~\bibnamefont
  {Habermann}}\ and\ \bibinfo {author} {\bibfnamefont {M.}~\bibnamefont
  {Mikhasenko}},\ }\href {https://doi.org/10.5281/zenodo.13741241} {\bibinfo
  {title} {Kaihabermann/decayangle: v1.1.2}} (\bibinfo {year} {2024}),\
  \bibinfo {note} {doi: 10.5281/zenodo.13741241}\BibitemShut {NoStop}%
\bibitem [{\citenamefont {Mikhasenko}(2024)}]{mikhasenko_2024_13713312}%
  \BibitemOpen
  \bibfield  {author} {\bibinfo {author} {\bibfnamefont {M.}~\bibnamefont
  {Mikhasenko}},\ }\href {https://doi.org/10.5281/zenodo.13713312} {\bibinfo
  {title} {Threebodydecays.jl: v0.12.2}} (\bibinfo {year} {2024}),\ \bibinfo
  {note} {doi: 10.5281/zenodo.13713312}\BibitemShut {NoStop}%
\end{thebibliography}%

\clearpage
\pagebreak

\appendix

\section{Matrices for Lorentz Transformations}\label{sec:matrices-for-lorentz-transformations}
In this appendix, we provide the explicit representation of the active transformation matrices for the \SOtreeone and \SLtwoC groups.

The \SLtwoC matrices are used to represent Lorentz transformations in spinor space.
The boost along the z~axis is given by:
\begin{align}
    B_z^{2\times2}(\gamma) = \cosh(\xi / 2) \cdot \mathds{1} + \sinh(\xi / 2) \cdot \sigma_z
\end{align}
where $\xi$ is the rapidity, related to the Lorentz factor $\gamma$ by $\xi = \cosh^{-1}(\gamma)$.

The rotation matrices for rotations around the y~axis and z~axis can be written as:
\begin{align}
    R_y^{2\times2}(\theta) = \cos(\theta / 2) \cdot \mathds{1} - i \sin(\theta / 2) \cdot \sigma_y \, , \\
    R_z^{2\times2}(\phi) = \cos(\phi / 2) \cdot \mathds{1} - i \sin(\phi / 2) \cdot \sigma_z \, .
\end{align}

The Pauli matrices $\sigma_y$ and $\sigma_z$ are defined as:
\begin{align}
    \sigma_y = \begin{bmatrix} 0 & -i \\ i & 0 \end{bmatrix}, \quad
    \sigma_z = \begin{bmatrix} 1 & 0 \\ 0 & -1 \end{bmatrix}
\end{align}

The \SOtreeone matrices represent Lorentz transformations in four-dimensional spacetime.
The matrices are used to transform the four-vectors of particles.
The boost along the z~axis can be written as
\begin{align}
    B_z^{4\times4}(\gamma) = \begin{bmatrix}
                                 \cosh(\xi) & 0 & 0 & \sinh(\xi) \\
                                 0          & 1 & 0 & 0          \\
                                 0          & 0 & 1 & 0          \\
                                 \sinh(\xi) & 0 & 0 & \cosh(\xi)
                             \end{bmatrix}\,,
\end{align}
where the elements of the matrix are indexed according to $p=(E,p_x,p_y,p_z)$ representation.

The rotation matrices around the y~axis and z~axis are defined as
\begin{align}
    R_y^{4\times4}(\theta) = \begin{bmatrix}
                                 1 & 0             & 0 & 0            \\
                                 0 & \cos(\theta)  & 0 & \sin(\theta) \\
                                 0 & 0             & 1 & 0            \\
                                 0 & -\sin(\theta) & 0 & \cos(\theta)
                             \end{bmatrix} \\
    R_z^{4\times4}(\phi) = \begin{bmatrix}
                               1 & 0          & 0           & 0 \\
                               0 & \cos(\phi) & -\sin(\phi) & 0 \\
                               0 & \sin(\phi) & \cos(\phi)  & 0 \\
                               0 & 0          & 0           & 1 \\
                           \end{bmatrix}
\end{align}

\section{Decoding parameters of Lorentz transformation}
\label{sec:decoding-parameters-of-lorentz-transformation}

This appendix provides the exact equations to obtain the parameters of a Lorentz transformation composed of an arbitrary set of boosts and rotations,
\begin{equation}
    \Lambda = O O \ldots O\,,
\end{equation}
where $O$ represents a boost or rotation operator. The objective is to convert this expression into a set of six parameters as defined in Eq.~\eqref{eq:arb.lorentz}.
We use two representations of the Lorentz group: \SOtreeone for intuitive decoding of angles and \SLtwoC for sensitivity to the $2\pi$ azimuthal phase.

First, apply the transformation $\Lambda$ to a vector at rest, $p_\rftext = (\vec 0, m)$. 
\begin{align*}
    p & = \Lambda \, p_\rftext                                                                                   \\
      & = R_z(\phi) R_y(\theta) B_z(\gamma) R_z(\psi_\rftext) R_y(\theta_\rftext) R_z(\phi_\rftext) \, p_\rftext \\
      & = R_z(\phi) R_y(\theta) B_z(\gamma) \, p_\rftext
\end{align*}
The first three rotations do not affect a vector with a zero momentum component. Thus, only the two rotations after the boost and the boost affect the vector. One can find the rapidity $\gamma$ and the angles $\theta$ and $\phi$ via:
\begin{align}
    \gamma & = \frac{E}{m} \, ,                                  \\
    \phi   & = \tan^{-1}(p_y, p_x)\, ,                           \\
    \theta & = \cos^{-1}\left(\frac{p_z}{|\vec{p}\,|}\right)\, ,
\end{align}
where $|\vec{p}\,|$ is the magnitude of the three-momentum vector and $E$ is the energy of the particle.

With the values of $\gamma$, $\theta$, and $\phi$ determined, one can extract the pure rotation part of $\Lambda$, denoted $R_\rftext$, by inverting the boost and post-boost rotations.

\begin{align}
    R_\rftext & = B_z^{-1}(\gamma) R_y^{-1}(\theta) R_z^{-1}(\phi) \,\Lambda
\end{align}
The rest-frame rotation angles can be calculated from the matrix elements $r_{ij}$ of $R_\rftext^{4 \times 4}$.
\begin{align}
    \phi_\rftext'  & = \tan^{-1}\left(r_{y,z}, r_{x,z}\right) \, ,  \\
    \psi_\rftext   & = \tan^{-1}\left(r_{z,y}, -r_{z,x}\right) \, , \\
    \theta_\rftext & = \cos^{-1}\left(r_{z,z}\right) \, .
\end{align}
With these angles one constructs the matrix \mbox{$\Lambda_{\mathrm{2\times 2}}^\rftext=R^{2\times2}(\phi_\rftext,\theta_\rftext,\psi_\rftext)$} as the \SLtwoC matrix. To acquire sensitivity to $2\pi$ jumps one can use the properties of the \SLtwoC representation and compare the reconstructed matrix to the original matrix.
\begin{align}
    \phi_\rftext =
    \begin{cases}
        \phi_\rftext'        & \text{if } \Lambda_{\mathrm{2\times 2}} =  \Lambda_{\mathrm{2\times 2}}^\rftext  \\
        \phi_\rftext' + 2\pi & \text{if } \Lambda_{\mathrm{2\times 2}} =  -\Lambda_{\mathrm{2\times 2}}^\rftext
    \end{cases}
\end{align}

\section{Example amplitude explicitly} \label{sec:example-amplitude-explicitly}

In this section,
we write down an expression for the amplitude of a four-body decay shown in Fig.~\ref{fig:wigner.rotation}.
The state of the total system is labeled by index $0$, its spin is $j_0$, and its spin projection along the $z$~axis is $\lambda_0$.
Particles in the final state are labeled with indices $1$ to $4$, with their respective spins and helicities denoted by $j_i$ and $\lambda_i$ ($i=1,\dots,4$). Note that the helicity values $\lambda_i$ are not necessarily equal to those in the center-of-momentum frame. The caveat on the topology-dependent definition of helicity is discussed in Sec.~\ref{sec:parametrization-of-cascade-reactions}.
Two decay topologies, $((12)3)4$ and $(1(23))4$, contribute to the amplitude expression.
The first topology, $((12)3)4$, is used as the reference and determines the frames in which the helicity values are defined. The amplitude for the second topology, $(1(23))4$, must align its quantization axes with those of the reference topology. The amplitude reads:
\begin{align*}
     & \mathcal{A}_{\lambda_0;\lambda_1,\lambda_2,\lambda_3,\lambda_4} = \mathcal{A}_{\lambda_0;\lambda_1,\lambda_2,\lambda_3,\lambda_4}^{\reftext}             \\
     & \quad +\sum_{\lambda_1',\lambda_2',\lambda_3',\lambda_4'} \mathcal{A}_{\lambda_0;\lambda_1',\lambda_2',\lambda_3',\lambda_4'}^{\chainind}                \\
     & \quad \qquad \times D^{j_1*}_{\lambda_1',\lambda_1}(\phi_{\chainind(\reftext)}^{1}, \theta_{\chainind(\reftext)}^{1}, \psi_{\chainind(\reftext)}^{1})    \\
     & \quad \qquad \times D^{j_2*}_{\lambda_2',\lambda_2}(\phi_{\chainind(\reftext)}^{2}, \theta_{\chainind(\reftext)}^{2}, \psi_{\chainind(\reftext)}^{2})    \\
     & \quad \qquad \times D^{j_3*}_{\lambda_3',\lambda_3}(\phi_{\chainind(\reftext)}^{3}, \theta_{\chainind(\reftext)}^{3}, \psi_{\chainind(\reftext)}^{3})    \\
     & \quad \qquad \times D^{j_4*}_{\lambda_4',\lambda_4}(\phi_{\chainind(\reftext)}^{4}, \theta_{\chainind(\reftext)}^{4}, \psi_{\chainind(\reftext)}^{4})\,, \\
\end{align*}
where $\mathcal{A}_{\lambda_0;\lambda_1,\lambda_2,\lambda_3,\lambda_4}$ denotes the full amplitude.
$\mathcal{A}_{\lambda_0;\lambda_1,\lambda_2,\lambda_3,\lambda_4}^{\reftext}$ and $\mathcal{A}_{\lambda_0;\lambda_1,\lambda_2,\lambda_3,\lambda_4}^{\chainind}$ represent the decay amplitudes for individual chains,
with $\reftext = ((12)3)4$, and $\chainind = (1(23))4$.
The second topology is included as a linear combination with coefficients determined by Wigner $D$-functions, which depend on kinematic angles. The angles $(\phi_{\chainind(\reftext)}^{i}, \theta_{\chainind(\reftext)}^{i}, \psi_{\chainind(\reftext)}^{i})$ are uniquely defined for each particle ($i = 1, \dots, 4$) by Eq.~\eqref{eq:L.rot}.
The algorithm for computing these angles is presented in Sec.~\ref{sec:parametrization-of-cascade-reactions}.

Amplitudes for both chains are computed using Eq.~\ref{eq:A.def}.
The first one reads:
\newpage
\begin{align*}
     & \mathcal{A}_{\lambda_0;\lambda_1,\lambda_2,\lambda_3,\lambda_4}^{\reftext} = \sum_{\eta=-j_{(12)3}}^{j_{(12)3}}\,\,\sum_{\rho=-j_{12}}^{j_{12}}                 \\
     & \quad D_{\lambda_0,\eta-\lambda_{4}}^{j_0*}(\phi^{\reftext}_{((12)3)4}, \theta^{\reftext}_{((12)3)4}, 0)\,H_{\eta,\lambda_{4}}^{\intdecgen{((12)3)4}{(12)3, 4}} \\
     & \quad \times D_{\eta,\rho-\lambda_{3}}^{j_{(12)3}*}(\phi^{\reftext}_{(12)3}, \theta^{\reftext}_{(12)3}, 0)\,H_{\rho,\lambda_{3}}^{\intdecgen{(12)3}{12, 3}}     \\
     & \quad \times D_{\rho,\lambda_{1}-\lambda_{2}}^{j_{12}*}(\phi^{\reftext}_{12}, \theta^{\reftext}_{12}, 0)\,H_{\lambda_1,\lambda_2}^{\intdecgen{12}{1,2}}\,.
\end{align*}
Arguments of the rotation matrices are helicity angles, $(\phi^\chainind_\nodeind, \theta^\chainind_\nodeind)$. These angles depend on the node of the topology where they are computed, and therefore carry the topology index $\chainind$ and the node specification $\nodeind$. For example, $(\phi^{\reftext}_{(12)3}, \theta^{\reftext}_{(12)3})$ denotes the spherical angles of momentum $(12)$ in the $123$ rest frame within the topology $\reftext=((12)3)4$. The summed indices $\eta$ and $\rho$ represent the helicities of the intermediate states, $(12)3$ and $12$, respectively. The helicity couplings $H$ are subject to the modeling, as discussed in Sec.~\ref{sec:introduction} and Sec.~\ref{sec:parametrization-of-cascade-reactions}.

The second amplitude is given by:
\begin{align*}
     & \mathcal{A}_{\lambda_0;\lambda_1,\lambda_2,\lambda_3,\lambda_4}^{\chainind} = \sum_{\eta=-j_{1(23)}}^{j_{1(23)}}\,\,\sum_{\rho=-j_{23}}^{j_{23}}            \\
     & \quad D_{\lambda_0,\eta-\lambda_4}^{j_0*}(\phi^\chainind_{(1(23))4}, \theta^\chainind_{(1(23))4}, 0)\,H_{\eta,\lambda_{4}}^{\intdecgen{(1(23))4}{1(23), 4}} \\
     & \quad \times D_{\eta,\lambda_{1}-\rho}^{j_{1(23)}*}(\phi^\chainind_{1(23)}, \theta^\chainind_{1(23)}, 0)\,H_{\lambda_{1},\rho}^{\intdecgen{1(23)}{1,(23)}}  \\
     & \quad \times D_{\rho,\lambda_{2}-\lambda_{3}}^{j_{23}*}(\phi^\chainind_{23}, \theta^\chainind_{23}, 0)\,H_{\lambda_2,\lambda_3}^{\intdecgen{23}{2,3}}\,,
\end{align*}
where the summed indices $\eta$ and $\rho$ denote helicity of the intermediate states, $1(23)$, and $23$.

\end{document}